\documentclass{emulateapj}
\usepackage{amsmath}

\newcommand{\grba}{GRB\,090323}
\newcommand{\grbb}{GRB\,090328}
\newcommand{\grbc}{GRB\,090902B}
\newcommand{\grbd}{GRB\,090926A}

\newcommand{\fermi}{{\em Fermi}}
\newcommand{\swift}{{\em Swift}}

\begin{document}

%%%%%%%%%%%%%%%%%%%%%%%%%%%%%%%%%%%%%%%%%%%%%%%%%%%%%%%%%%%%%%%%%%%%%%%%%
\shorttitle{Afterglow Observations of \fermi-LAT GRBs}
\shortauthors{Cenko et al.}

%%%%%%%%%%%%%%%%%%%%%%%%%%%%%%%%%%%%%%%%%%%%%%%%%%%%%%%%%%%%%%%%%%%%%%%%%
\title{Afterglow Observations of \fermi-LAT Gamma-Ray Bursts and the 
       Emerging Class of Hyper-Energetic Events}

\author{S.~B.~Cenko\altaffilmark{1}, D.~A.~Frail\altaffilmark{2},
        F.~A.~Harrison\altaffilmark{3}, J.~B.~Haislip\altaffilmark{4},
        D.~E.~Reichart\altaffilmark{4}, N.~R.~Butler\altaffilmark{1,5},
        B.~E.~Cobb\altaffilmark{1}, A.~Cucchiara\altaffilmark{6},
        E.~Berger\altaffilmark{7}, J.~S.~Bloom\altaffilmark{1},
        P.~Chandra\altaffilmark{8}, D.~B.~Fox\altaffilmark{6},
        D.~A.~Perley\altaffilmark{1}, J.~X.~Prochaska\altaffilmark{9},
        A.~V.~Filippenko\altaffilmark{1}, K.~Glazebrook\altaffilmark{10},
        K.~M.~Ivarsen\altaffilmark{4},
        M.~M.~Kasliwal\altaffilmark{11}, S.~R.~Kulkarni\altaffilmark{11},
        A.~P.~LaCluyze\altaffilmark{4}
        S.~Lopez\altaffilmark{12}, 
        A.~N.~Morgan\altaffilmark{1}, M.~Pettini\altaffilmark{13,14},
        and V.~R.~Rana\altaffilmark{3}
        }

\altaffiltext{1}{Department of Astronomy,
  University of California, Berkeley, CA 94720-3411, USA.}
\altaffiltext{2}{National Radio Astronomy Observatory, 1003 Lopezville
  Road, Socorro, NM 87801, USA.}
\altaffiltext{3}{Space Radiation Laboratory, California Institute of
  Technology, MS 105-24, Pasadena, CA 91125, USA.}
\altaffiltext{4}{Department of Physics and Astronomy, University of
  North Carolina, Chapel Hill, NC 27599, USA.}
\altaffiltext{5}{Einstein Fellow.}
\altaffiltext{6}{Department of Astronomy and Astrophysics, Pennsylvania
  State University, 525 Davey Laboratory, University Park, PA 16802, USA.}
\altaffiltext{7}{Harvard-Smithsonian Center for Astrophysics, 60 Garden
  Street, Cambridge, MA 02138, USA.}
\altaffiltext{8}{Department of Physics, Royal Military College of Canada,
  Kingston, ON, Canada.}
\altaffiltext{9}{Department of Astronomy and Astrophysics, UCO/Lick
  Observatory, University of California, 1156 High Street, Santa Cruz,
  CA 95064, USA.}
\altaffiltext{10}{Centre for Astrophysics and Supercomputing, Swinburne
  University of Technology, 1 Alfred St, Hawthorn, Victoria 3122,
  Australia.}
\altaffiltext{11}{Department of Astronomy, California Institute of
  Technology, MS 105-24, Pasadena, CA 91125, USA.}
\altaffiltext{12}{Departamento de Astronom\'ia, Universidad de Chile,
  Casilla 36-D, Santiago, Chile.}
\altaffiltext{13}{Institute of Astronomy, Madingley Road, Cambridge CB3 0HA,
  UK.}
\altaffiltext{14}{International Center for Radio Astronomy Research,
  University of Western Australia, 35 Stirling Hwy, Crawley,
  WA 6009, Australia.}

%%%%%%%%%%%%%%%%%%%%%%%%%%%%%%%%%%%%%%%%%%%%%%%%%%%%%%%%%%%%%%%%%%%%%%%%%%%

\slugcomment{Submitted to ApJ}

%%%%%%%%%%%%%%%%%%%%%%%%%%%%%%%%%%%%%%%%%%%%%%%%%%%%%%%%%%%%%%%%%%%%%%%%%%%
\begin{abstract}
  We present broadband (radio, optical, and X-ray) light curves and
  spectra of the afterglows of four long-duration gamma-ray bursts 
  (GRBs\,090323, 090328, 090902B, and 090926A) detected by the
  Gamma-Ray Burst Monitor (GBM) and Large Area Telescope (LAT)
  instruments on the \fermi\ satellite. With its wide spectral
  bandpass, extending to GeV energies, \fermi\ is sensitive to GRBs
  with very large isotropic energy releases (10$^{54}$ erg). Although
  rare, these events are particularly important for testing GRB
  central-engine models. When combined with spectroscopic redshifts,
  our afterglow data for these four events are able to constrain jet
  collimation angles, the density structure of the circumburst medium,
  and both the true radiated energy release and the kinetic energy of
  the outflows. In agreement with our earlier work, we find that the
  relativistic energy budget of at least one of these events 
  (GRB\,090926A) exceeds the canonical value of $10^{51}$\,erg by
  an order of magnitude. Such energies pose a severe challenge for
  models in which the GRB is powered by a magnetar or neutrino-driven
  collapsar, but remain compatible with theoretical expectations for
  magneto-hydrodynamical collapsar models (e.g., the 
  Blandford-Znajek mechanism).  Our jet opening angles ($\theta$) are 
  similar to those found for pre-\fermi\ GRBs, but the large initial 
  Lorentz factors ($\Gamma_{0}$) inferred from the detection of GeV photons 
  imply $\theta\Gamma_{0} \approx 70$--90, values which are above
  those predicted in magnetohydrodynamic models of jet acceleration. Finally, 
  we find that these \fermi-LAT events preferentially occur in a low-density
  circumburst environment, and we speculate that this might result
  from the lower mass-loss rates of their lower-metallicity
  progenitor stars.  Future studies of \fermi-LAT afterglows in the radio
  with the order-of-magnitude improvement in sensitivity offered by
  the Extended Very Large Array should definitively establish the relativistic 
  energy budgets of these events.
\end{abstract}
%%%%%%%%%%%%%%%%%%%%%%%%%%%%%%%%%%%%%%%%%%%%%%%%%%%%%%%%%%%%%%%%%%%%%%%%%%%%

%%%%%%%%%%%%%%%%%%%%%%%%%%%%%%%%%%%%%%%%%%%%%%%%%%%%%%%%%%%%%%%%%%%%%%%%%%%%

%%%%%%%%%%%%%%%%%%%%%%%%%%%%%%%%%%%%%%%%%%%%%%%%%%%%%%%%%%%%%%%%%%%%%%%%%%%
\section{Introduction}
%%%%%%%%%%%%%%%%%%%%%%%%%%%%%%%%%%%%%%%%%%%%%%%%%%%%%%%%%%%%%%%%%%%%%%%%%%%
\label{sec:intro}
Long-duration gamma-ray bursts (GRBs\footnote{Throughout this work,
we shall refer exclusively to long-duration GRBs (i.e., those 
apparently with massive-star progenitors) unless explicitly stated otherwise.}),
like hydrogen-deficient Type Ib/c supernovae (SNe Ib/c), result
from the gravitational collapse of the evolved core of a massive
star.  The main characteristic that sets GRBs apart from other SNe 
is that a substantial fraction of the energy of the explosion is
coupled to relativistic ejecta. A compact central engine is
responsible for accelerating and collimating these jet-like outflows 
and driving the SN explosions \citep{wb06,grf09,scp+10}.  The 
precise nature of the central engine which powers GRB-SNe, however, 
remains an open question.

Motivated by empirical constraints, all viable central-engine models 
for long-duration GRBs share some common characteristics
(e.g., \citealt{p05}).  They must produce a collimated outflow with an
initial Lorentz factor ($\Gamma_{0}$) of a few hundred on observed 
time scales of 10--100\,s, with luminosities and kinetic energies of order 
10$^{50}$\,erg s$^{-1}$ and 10$^{51}$ erg, respectively. Leading models 
include the ``collapsar'' model in which a relativistic jet is produced 
from a rotating black hole/accretion disk system \citep{w93,mw99}, and the 
``magnetar'' model in which the rapid energy loss from a newly born 
millisecond neutron star (formed either from the gravitational collapse of a 
massive star or from accreting or coalescing white dwarfs) with a 
10$^{15}$\,G magnetic field drives a Poynting flux-dominated relativistic 
outflow \citep{u92}.

These and other more exotic models for GRB central engines are
highly constrained by their energetics. The prompt high-energy
emission, when combined with a spectroscopically determined redshift
(and hence distance measurement), yields the isotropic radiated 
gamma-ray energy ($E_{\gamma,\mathrm{iso}}$). Well-sampled afterglow 
observations allow both a measurement of the degree of collimation
(and hence the true beaming-corrected energy release in the prompt
emission, $E_\gamma$) and the kinetic energy remaining in the
shock that powers the broadband afterglow emission ($E_{\mathrm{KE}}$). 
Such measurements, made nearly a decade ago, pointed to a total {\it
 relativistic} energy yield ($E_{\mathrm{rel}} \approx E_\gamma +
E_{\mathrm{KE}}$) of $\sim 10^{51}$\,erg
\citep{fks+01,pk01b,fw01,bfk03,bkf03}.

Since that time, there has been growing evidence for a considerable
range in the relativistic energy scale $E_{\mathrm{rel}}$, suggesting either a
diversity in central engines or their properties. Most notably, a
population of nearby (redshift $z \lesssim 0.1$) subenergetic 
long-duration GRBs have been identified \citep{bfk03,skb+04,skn+06}.
They too are associated with SNe~Ib/c, but their relativistic energy 
release is a factor of 100 less than that of typical cosmological GRBs 
and their outflows are significantly less collimated (quasi-spherical).  
Since they can only be detected at low redshifts where the comparative
volume for discovery is low, they are small in total number.  But their 
volumetric rate is inferred to be 10--100 times larger than that of the
more distant long-duration GRBs \citep{skn+06,cbv+06,lzv+07}.

More recently, evidence has been growing for a class of
GRBs whose total relativistic energy release is at least an order of 
magnitude above the canonical value of $10^{51}$\,erg (e.g.,
\citealt{cfh+10} and references therein).  Unlike subluminous events,
the total energy budget of these hyper-energetic events poses a 
significant challenge for some progenitor models.  In particular, models
in which the GRB is powered by a magnetar or a neutrino-driven
collapsar are strongly disfavored.  On the other hand, collapsars
driven by magneto-hydrodynamical (MHD) processes, such as the 
Blandford-Znajek mechanism \citep{bz77}, can naturally accomodate
energy budgets as large as $10^{53}$\,erg.

Unfortunately, it has been rather difficult
to constrain the beaming-corrected energetics for the hundreds
of GRBs detected by the \swift\ satellite \citep{gcg+04}.  The reasons 
for this difficulty are now largely understood.
First, the relatively narrow energy bandpass (15--150\,keV) can miss
entirely the peak of the gamma-ray spectrum, making estimates of
$E_{\gamma,\mathrm{iso}}$ highly uncertain.
Second, there has been a dearth of measurements of jet opening angles
(e.g., \citealt{p07,kb08,lrz+08,rlb+09}) and well-sampled
multi-wavelength GRB afterglows (used to derived the afterglow kinetic
energy $E_{\mathrm{KE}}$). \swift\ GRBs are on average more than twice
as distant \citep{jlf+06} and therefore significantly fainter ($\sim
1.5$\,mag in the optical; \citealt{bkf+05,kkz+07}) than GRBs in previous
samples.  In large part this is due to selection effects: a
combination of bandpass and sensitivity from \swift\ has
preferentially selected the faint end of the luminosity function --- GRBs
with low isotropic energy release but large opening angles
\citep{psf03}.

\begin{figure}[t]
\epsscale{1.2}
\plotone{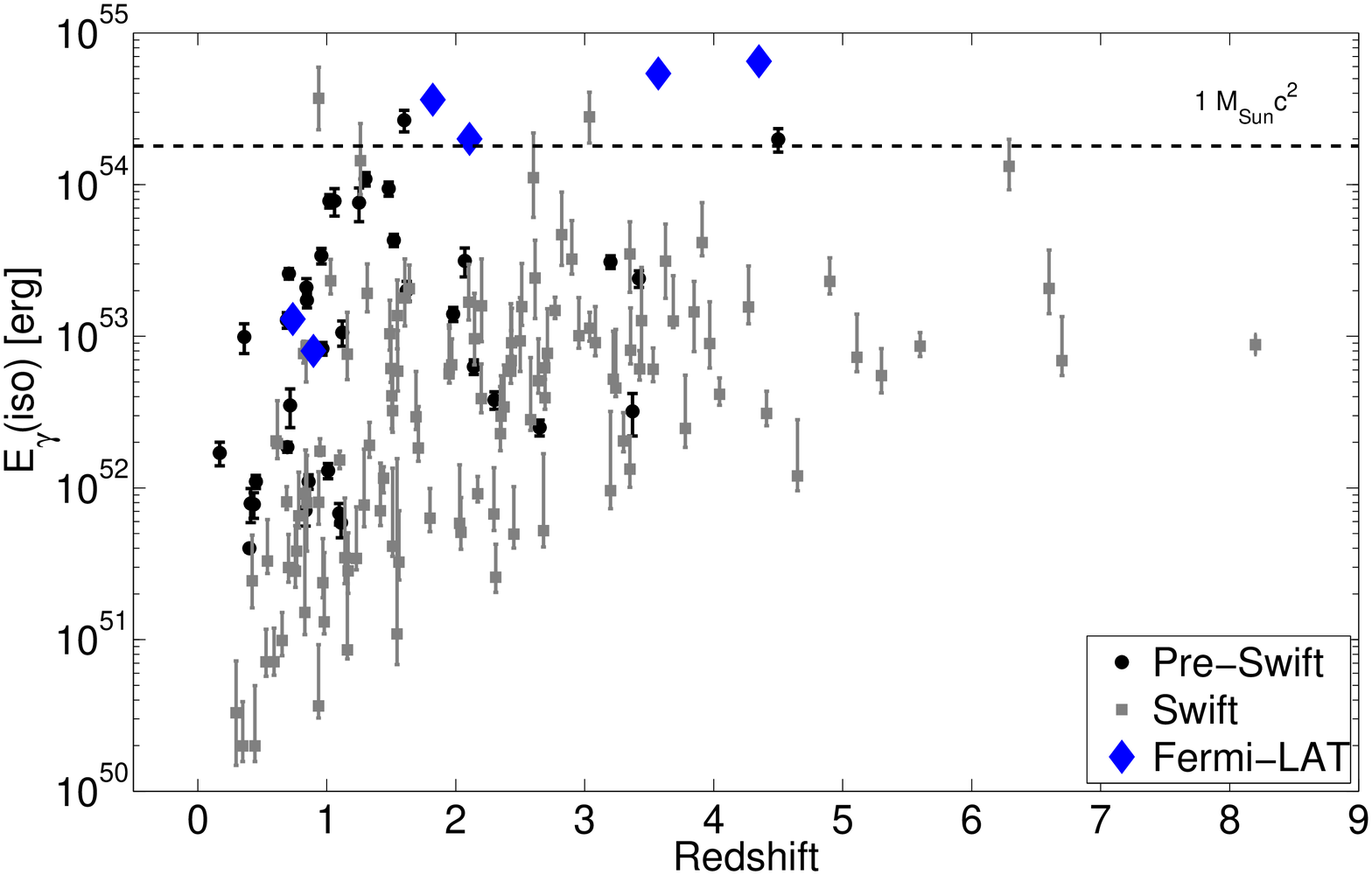}
%\plotone{../Figures/egamma_lat.eps}
\caption{Prompt isotropic gamma-ray energy release ($E_{\gamma,\mathrm{iso}}$)
of GRBs.  With its soft, narrow bandpass (15--150\,keV), \swift\
typically selects events with smaller isotropic energy release but larger
opening angles than previous missions, which triggered predominantly in the
MeV bandpass \citep{psf03}.  GRBs detected at GeV energies with the 
\fermi-LAT all fall at the brightest end of the isotropic energy distribution,
and must therefore be highly collimated to achieve a canonical 
beaming-corrected energy release of $\sim 10^{51}$\,erg.  References:
pre-\swift: \citet{a06}; \swift: \citet{bkb+07}; \fermi-LAT: 
\citet{gck+09}, this work.}
\label{fig:egamma}
\end{figure}

With its nearly seven decades in energy coverage (10\,keV
-- 100\,GeV), \fermi\ can provide unparalleled constraints on this 
subsample of the most luminous events. In light of the empirical relation 
between the peak energy of the gamma-ray spectrum and the isotropic 
gamma-ray energy release (the $E_{\mathrm{p}}$--$E_{\gamma,\mathrm{iso}}$, 
or ``Amati'' relation; \citealt{a06}), MeV/GeV events detected by either 
the Gamma-Ray Burst Monitor (GBM, 8\,keV -- 40\,MeV; \citealt{mlb+09}) or 
the Large Area Telescope (LAT, 20\,MeV -- 300\,GeV; \citealt{aaa+09c}) 
onboard \fermi\ preferentially select a sample of GRBs with large isotropic
energy release (Figure~\ref{fig:egamma}). High-$E_{\gamma,\mathrm{iso}}$ 
events also have brighter X-ray and optical afterglows on average 
\citep{nfp09}.  Follow-up afterglow observations can then determine whether 
these GRBs are highly beamed events ($\theta \lesssim 2^{\circ}$) with 
a typical energy release or true hyper-energetic GRBs.

%%%%%%%%%%%%%%%%%%%%%%%%%%%%%%%%%%%%%%%%%%%%%%%%%%%%%%%%%%%%%
%%%%%%%%%%%%%%%%%%%%%%%%%%%%%%%%%%%%%%%%%%%%%%%%%%%%%%%%%%%%%

The \fermi-LAT offers a further advantage over previous GRB missions
sensitive only at MeV and keV energies by providing strict constraints
on the initial Lorentz factor of the relativistic outflow.  To avoid 
$e^{+}-e^{-}$ pair production (and the accompanying thermal spectrum),
the GRB jet must be moving towards the observer with ultra-relativistic 
speeds (the ``compactness'' problem; \citealt{cr78}).  The higher the 
energy of the most energetic photon detected from a GRB, the more
strict the lower limit on the outflow Lorentz factor will be.
Combining the Lorentz factor limits for the most relativistic GRBs with 
inferred jet opening angles from broadband afterglow models can provide
critical diagnostics of the jet acceleration mechanism.

%%%%%%%%%%%%%%%%%%%%%%%%%%%%%%%%%%%%%%%%%%%%%%%%%%%%%%%%%%%%%
\begin{figure*}[t]
\plotone{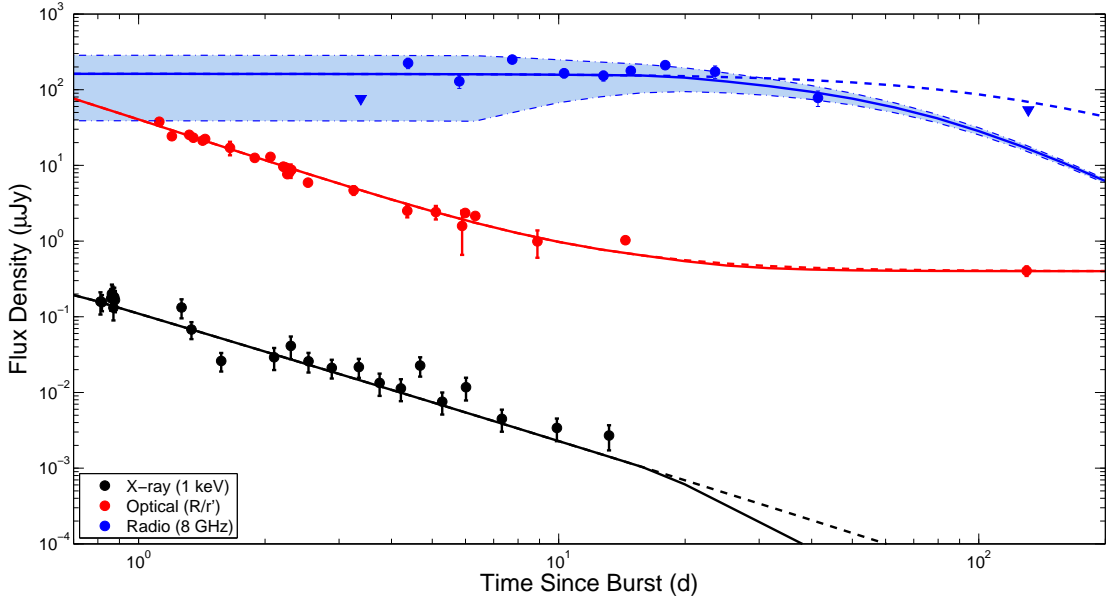}
%\plotone{../Figures/0323.eps}
\caption{The broadband radio (blue), optical (red), and X-ray (black) light
curve of \grba.  The best-fit model is plotted in solid lines (see 
Table~\ref{tab:0323mod} for parameters).  The identical model 
parameters for an isotropic
explosion are plotted as the dashed lines.  The strength of the possible
modulation of the radio afterglow caused by interstellar scintillation
(e.g., \citealt{fwk00}) is indicated by the light-blue shaded region.  
The model provides a reasonable
fit in all bandpasses.  It is clear that any jet break must occur at $t > 
10$\,days, although the upper bound on the jet break time is only weakly 
constrained.}
\label{fig:0323}
\end{figure*}
%%%%%%%%%%%%%%%%%%%%%%%%%%%%%%%%%%%%%%%%%%%%%%%%%%%%%%%%%%%

Here we report on broadband (radio, optical, and X-ray) observations
of four long-duration GRBs detected by the \fermi-LAT at GeV energies:
GRBs\,090323, 090328, 090902B, and 090926A.  For each event we construct
afterglow models to constrain the collimation and beaming-corrected
energetics, and we compare these LAT events with previous GRBs detected
at other energies (i.e., keV energies from \swift, and MeV energies from 
pre-\swift\ satellites).  For three of these GRBs, we also present
the optical spectra used to determine the afterglow redshift.  A more
thorough analysis of the host-galaxy properties of these events will
be presented in a forthcoming work.

Throughout this paper, we adopt a standard $\Lambda$CDM cosmology with
$H_{0}$ = 71\,km s$^{-1}$ Mpc$^{-1}$, $\Omega_{\mathrm{m}} = 0.27$, and
$\Omega_{\Lambda} = 1 - \Omega_{\mathrm{m}} = 0.73$ \citep{sbd+07}.  We 
define the flux-density power-law temporal and spectral decay indices 
$\alpha$ and $\beta$ as $f_{\nu} \propto t^{-\alpha} \nu^{-\beta}$ (e.g., 
\citealp{spn98}).  All quoted uncertainties are 1$\sigma$ ($68\%$) confidence 
intervals unless otherwise noted.

%%%%%%%%%%%%%%%%%%%%%%%%%%%%%%%%%%%%%%%%%%%%%%%%%%%%%%%%%%%%%%%%%%%%%%
\section{Observations}
\label{sec:obs}
%%%%%%%%%%%%%%%%%%%%%%%%%%%%%%%%%%%%%%%%%%%%%%%%%%%%%%%%%%%%%%%%%%%%%%

%%%%%%%%%%%%%%%%%%%%%%%%%%%%%%%%%%%%%%%%%%%%%%%%%%%%%%%%%%%%%%%%%%%%%%
\subsection{\grba}
\label{sec:0323obs}
%%%%%%%%%%%%%%%%%%%%%%%%%%%%%%%%%%%%%%%%%%%%%%%%%%%%%%%%%%%%%%%%%%%%%%

%%%%%%%%%%%%%%%%%%%%%%%%%%%%%%%%%%%%%%%%%%%%%%%%%%%%%%%%%%%
\begin{figure*}[t]
\plotone{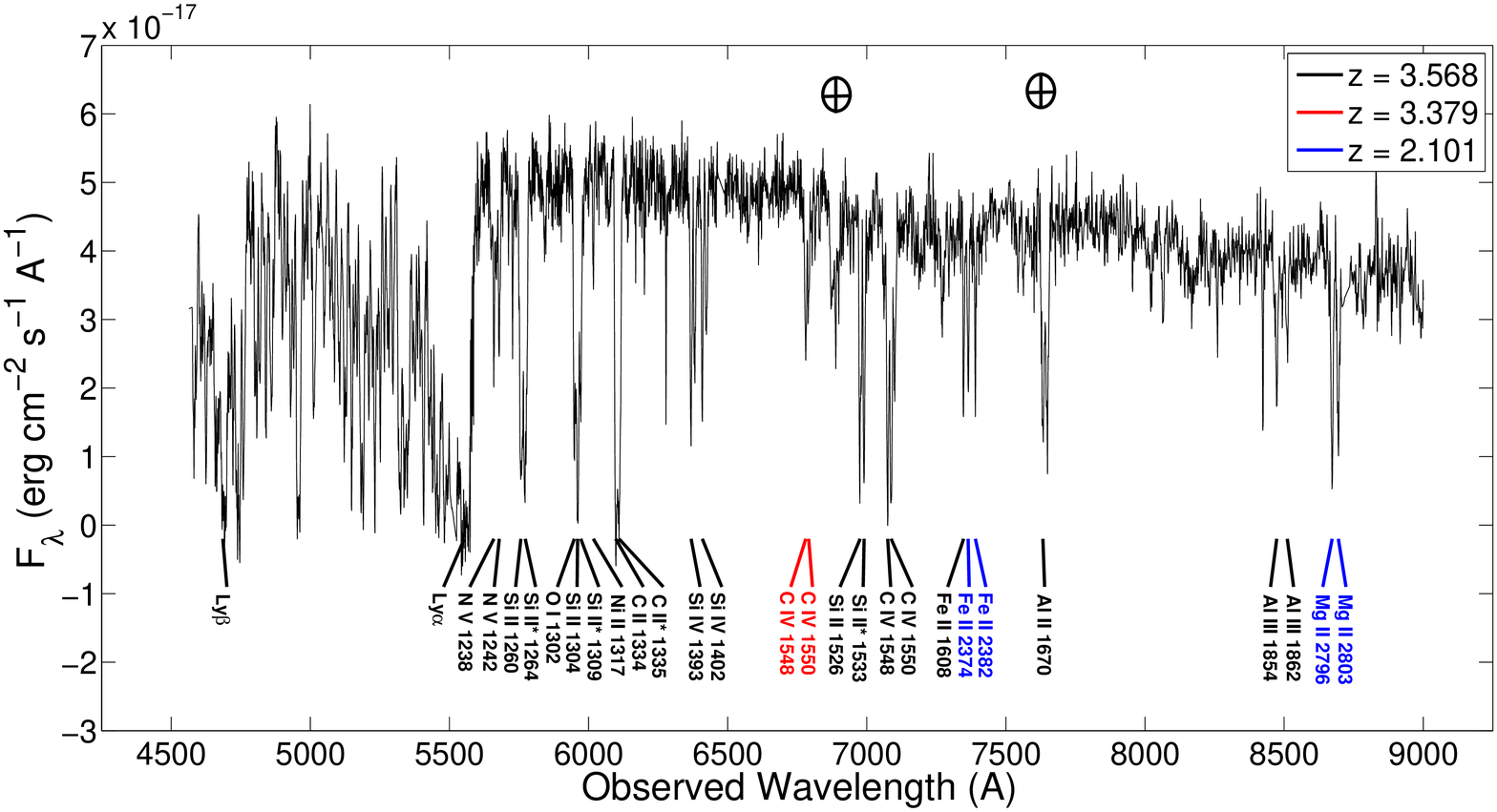}
%\plotone{../Figures/0323_spec.eps}
\caption{GMOS-S optical spectrum of the afterglow of \grba.  The broad 
absorption feature at $\lambda \approx 5570$\,\AA\ results from Ly$\alpha$
in the GRB host galaxy.  We identify a number of strong, narrow absorption
features redward of Ly$\alpha$ from the GRB host galaxy at a common redshift
of $3.568 \pm 0.004$ (black annotations).  The emission blueward of 
Ly$\alpha$ is strongly affected by the Ly$\alpha$ forest.  We identify two 
additional intervening absorbers, based on \ion{Mg}{2} $\lambda\lambda$2796, 
2803 ($z = 2.101$; blue annotations) and \ion{C}{4} $\lambda\lambda$1548, 
1550 ($z = 3.379$; red annotations).  Areas of the spectrum affected
by telluric absorption in the atmosphere of the Earth are marked with
crossed circles.  Because of the increased noise in the NIR due to
fringing, we have cut off the plot at $\lambda = 9000$\,\AA. }
\label{fig:0323spec}
\end{figure*}
%%%%%%%%%%%%%%%%%%%%%%%%%%%%%%%%%%%%%%%%%%%%%%%%%%%%%%%%%%%

%%%%%%%%%%%%%%%%%%%%%%%%%%%%%%%%%%%%%%%%%%%%%%%%%%%%%%%%%%%%%%%%%%%%%%
\subsubsection{High-Energy Properties}
\label{sec:0323high}
\grba\ was detected by the \fermi\ GBM at 00:02:42.63 on 23 March
2009 (\citealt{GCN.9021}; UT dates are used throughout this work).
In the 8\,keV to 40\,MeV bandpass of the GBM, the light curve
was multipeaked with a duration\footnote{It is customary to report GRB
durations measured as the time between the arrival of 5\% and 95\% of the
background-subtracted fluence.  This quantity is referred to as $t_{90}$.} 
of $t_{90} \approx 150$\,s.  \grba\ was also
detected at MeV energies by several of the satellites comprising
the Inter-Planetary Network (IPN; \citealt{hbk+99}), including the
Konus instrument on the \textit{Wind} satellite \citep{GCN.9023,GCN.9030}.

Examining only the first 70\,s of GBM data\footnote{After this point,
the \fermi\ spacecraft slewed to position the GRB at the center of the
LAT field of view, causing rapid changes in the GBM background level.},
\citet{GCN.9035} find that the prompt spectrum is well described by
a power law having an exponential cutoff at high 
energies with $\alpha = -0.89 \pm 0.03$ and 
$E_{\mathrm{p}} = 697 \pm 51$\,keV.  The resulting fluence in the 
8--10$^{3}$\,keV (observer frame) bandpass is $f_{\gamma} = (1.00 \pm 0.01) 
\times 10^{-4}$\,erg cm$^{-2}$.  The Konus-\textit{Wind} instrument was able 
to measure the high-energy spectrum over the entire duration of the MeV 
emission.  Fitting a Band model \citep{bmf+93} to their spectrum, 
\citet{GCN.9030}
find a reasonable fit with $\alpha = -0.96_{-0.09}^{+0.12}$, $\beta = 
-2.09_{-0.22}^{+0.16}$, $E_{\mathrm{p}} = 416_{-73}^{+76}$\,keV, and $f_{\gamma} = 
2.02_{-0.25}^{+0.28} \times 10^{-4}$\,erg\,cm$^{-2}$ (20--10$^{4}$\,keV
observer-frame bandpass).  

In order to facilitate direct comparisons between events, we must 
transform this high-energy fluence to a common (rest-frame) bandpass
(i.e., a ``k''-correction; \citealt{bfs01}).  Here we adopt the rest-frame
1--10$^{4}$\,keV bandpass, as this encompasses the full range of peak 
energies observed from GRBs \citep{bmf+93} without (for most pre-\fermi\
satellites) requiring large extrapolations outside the observed bandpass.
At $z = 3.568$ ($\S$~\ref{sec:0323spec}),
The Konus-\textit{Wind} measurement corresponds to a prompt fluence 
of $f_{\gamma} = (1.50 \pm 0.20)
\times 10^{-4}$\,erg\,cm$^{-2}$ in the 1--10$^{4}$\,keV rest-frame
bandpass.  Given the limited temporal coverage of the GBM spectrum,
we shall adopt this value for the prompt fluence of \grba\ for the 
remainder of this work.

In addition to the detection at MeV energies by the GBM, \grba\ was
also detected at GeV energies by the \fermi\ LAT.  Like several 
previous GRBs observed at GeV energies (e.g., 
\citealt{hdm+94,gdk+03,gmf+08,aaa+09d}), the GeV emission began
several seconds after the MeV emission is detected, and remains
significant long ($\sim 1000$\,s) after the MeV component has faded
beyond detectability \citep{GCN.9021}.  The highest energy photon
from the direction of \grba\ had an energy of $E \approx 7.5$\,GeV
\citep{p09}.  However, this photon arrived after the prompt MeV 
emission ended, and so its origin (i.e., prompt emission or afterglow,
or, in physical terms, internal or external shock-related)
is somewhat uncertain.  The highest energy photon 
detected by the LAT during the formal MeV $t_{90}$ measurement had an
energy of $E \approx 500$\,keV (at $t \approx 90$\,s; \citealt{p09}).

%%%%%%%%%%%%%%%%%%%%%%%%%%%%%%%%%%%%%%%%%%%%%%%%%%%%%%%%%%%%%%%%%%%%%%
\subsubsection{Afterglow Observations}
\label{sec:0323ag}
The \swift\ satellite began observations of the field 
of \grba\ with the onboard X-ray Telescope (XRT; \citealt{bhn+05})
at 19:27 on 23 March 2009 ($\sim 19.4$\,hours after the burst; 
\citealt{GCN.9024}).  A candidate afterglow was promptly identified 
at $\alpha = 12^{\mathrm{h}} 42^{\mathrm{m}} 50.^{\mathrm{s}}26$, $\delta=
+17^{\circ} 03\arcmin 14\farcs2$, with a 90\% containment radius of
$2\farcs7$ (J2000.0).  The XRT continued to monitor the evolution of
the fading counterpart for the following two weeks.  We have obtained
the XRT light curve from the online compilation of
N.R.B.\footnote{\texttt{http://astro.berkeley.edu/$\sim$nat/swift}; see 
\citet{bk07} for details.}; the resulting evolution is plotted in 
Figure~\ref{fig:0323}.

The optical afterglow of \grba\ was discovered shortly thereafter by 
GROND \citep{GCN.9026}.  The optical/NIR spectral energy distribution, 
constructed from simultaneous imaging in the 
$g^{\prime}r^{\prime}i^{\prime}z^{\prime}JHK$ filters, implied a spectral 
steepening around the observed $g^{\prime}$ band.  Associating this break
with absorption from Ly$\alpha$ in the GRB host galaxy, \citet{GCN.9026}
derive a photometric redshift of $4.0 \pm 0.3$ for \grba.

We began observations of the afterglow of \grba\ with the automated Palomar 
60\,inch (1.5\,m) telescope (P60; \citealt{cfm+06}) beginning on 2009 
March 24 \citep{GCN.9027}.  Images were obtained in the Sloan $r^{\prime}$ 
and $i^{\prime}$ filters and individual frames were automatically reduced 
using our custom IRAF\footnote{IRAF is distributed by the National Optical 
Astronomy Observatory, which is operated by the Association for Research in 
Astronomy, Inc., under cooperative agreement with the National Science 
Foundation.} software pipeline.  To increase the signal-to-noise ratio (S/N), 
individual frames were astrometrically aligned using the Scamp software 
package and coadded using Swarp\footnote{See 
\texttt{http://astromatic.iap.fr}.}.  We used aperture photometry to 
extract the flux of the afterglow from these coadded frames with the 
aperture radius roughly matched to the full width at half-maximum intensity 
(FWHM) of the point-spread function (PSF). Aperture magnitudes were then 
calibrated relative to field sources from the Sloan Digital Sky Survey Data 
Release 7 \citep{aaa+09e}.  Imaging continued on subsequent nights with P60 
until the afterglow was below our detection threshold.  The results of this 
monitoring campaign, uncorrected for foreground Galactic extinction 
($E[B-V] = 0.025$\,mag; \citealt{sfd98}), are presented in
Table~\ref{tab:0323opt}.

We obtained additional imaging of the field of \grba\ with the two Gemini
Multi-Object Spectrographs (GMOS-N and GMOS-S; \citealt{hja+04}).  Images 
were taken at both Gemini North and Gemini South in the Sloan $r^{\prime}$ 
and $i^{\prime}$ filters, and they were reduced using the IRAF \texttt{gemini} 
package.  Photometry was performed with the same methodology used for
the P60 imaging of the field, and the resulting measurements are shown in 
Table~\ref{tab:0323opt}.

Additionally, we have compiled optical and NIR measurements of the 
afterglow of \grba\ from the GCN\footnote{See 
\texttt{http://gcn.gsfc.nasa.gov/gcn3\_archive.html.}} Circulars and 
included these in Table~\ref{tab:0323opt}.  The majority of the late-time 
measurements (of most interest to us for modeling purposes) were obtained in 
the $R$ band.  We note that most of these measurements were calibrated
with respect to a single object (1070-0238439) from the USNO-B1 catalog 
\citep{mlc+03}, following \citet{GCN.9033}.  Using the SDSS observations of 
this field and the filter transformations of \citet{jga06}, we find an 
$R$-band magnitude for this source of $R = 17.31$\,mag (compared to an 
assumed value of 17.15\,mag from \citealt{GCN.9033}).  We therefore offset 
all reported $R$-band magnitudes from the GCN Circulars by $+0.16$\,mag.

Finally, we observed the afterglow of \grba\ with the Very
Large Array\footnote{The Very Large Array is operated by the National 
Radio Astronomy Observatory, a facility of the National Science 
Foundation operated under cooperative agreement by Associated 
Universities, Inc.} (VLA) beginning a few days after the initial burst 
trigger. The flux-density scale was tied to 3C\,286 or 3C\,147 and the 
phase was measured by switching between the GRB and a nearby bright 
point-source calibrator.  To maximize sensitivity, the full VLA continuum 
bandwidth (100 MHz) was recorded in two 50 MHz bands.  Data reduction was 
carried out following standard practice in the {\it AIPS} software package.

Our initial VLA detection of \grba\ was reported by \citet{GCN.9043}.  
\grba\ was also detected at radio wavelengths by the Westerbork
Synthesis Radio Telescope (WSRT; \citealt{GCN.9047}).  The full set of VLA 
measurements is listed in Table~\ref{tab:0323rad}.  In order to improve 
the S/N and to reduce the modulation of the light curve caused 
by interstellar scintillation (e.g., \citealt{fwk00}), we binned the 
data from adjacent epochs. These binned points were used for our
afterglow modeling (\S~\ref{sec:results}) and are  
plotted in Figure~\ref{fig:0323}.

%%%%%%%%%%%%%%%%%%%%%%%%%%%%%%%%%%%%%%%%%%%%%%%%%%%%%%%%%
\begin{deluxetable}{l r c c c}
\tablewidth{0pt}
\tablecaption{Radio Observations of GRB\,090323}
\tablehead{
\colhead{Date} & \colhead{$\Delta{t}$} &\colhead{$\nu$} &\colhead{$f_\nu$} 
  &\colhead{Facility}\\
\colhead{(UT)} & (days) & (GHz) & ($\mu$Jy) & 
}
\startdata
Mar 26.38 & 3.38 & 8.46 & 27 $\pm$ 38 & VLA \\
Mar 27.38 & 4.38 & 8.46 & 225 $\pm$ 35 & VLA \\
Mar 27.99 & 4.99 & 4.9 & 105 $\pm$ 24 & WSRT\tablenotemark{a} \\
Mar 28.43 & 5.43 & 8.46 & 100 $\pm$ 40 & VLA \\
Mar 29.16 & 6.16 & 8.46 & 157 $\pm$ 31 & VLA \\
Mar 30.18 & 7.18 & 8.46 & 219 $\pm$ 39 & VLA \\
Mar 31.32 & 8.32 & 8.46 & 281 $\pm$ 38 & VLA \\
Apr 1.30 & 9.30 & 8.46 & 164 $\pm$ 35 & VLA \\
Apr 3.29 & 11.29 & 8.46 & 166 $\pm$ 27 & VLA \\
Apr 3.28 & 11.28 & 4.86 & 110 $\pm$ 45 & VLA \\
Apr 4.41 & 12.41 & 8.46 & 183 $\pm$ 35 & VLA \\
Apr 5.14 & 13.14 & 8.46 & 123 $\pm$ 29 & VLA \\
Apr 6.28 & 14.28 & 8.46 & 312 $\pm$ 27 & VLA \\
Apr 7.42 & 15.42 & 8.46 & 43 $\pm$ 27 & VLA \\
Apr 9.44 & 17.44 & 8.46 & 127 $\pm$ 30 & VLA \\
Apr 10.42 & 18.42 & 8.46 & 295 $\pm$ 27 & VLA \\
Apr 14.09 & 22.09 & 8.46 & 178 $\pm$ 56 & VLA \\
Apr 17.04 & 25.04 & 8.46 & 167 $\pm$ 33 & VLA \\
Apr 25.46 & 33.46 & 8.46 & 78 $\pm$ 32 & VLA \\
May 3.41 & 41.41 & 8.46 & 77 $\pm$ 31 & VLA \\
May 11.35 & 49.35 & 8.46 & 61 $\pm$ 29 & VLA \\
Aug 1.18 & 131.18 & 8.46 & $-13 \pm 27$ & VLA
\enddata
\tablenotetext{a}{Reference: \citet{GCN.9047}.}
\label{tab:0323rad}
\end{deluxetable}
%%%%%%%%%%%%%%%%%%%%%%%%%%%%%%%%%%%%%%%%%%%%%%%%%%%%%%%%%%%

%%%%%%%%%%%%%%%%%%%%%%%%%%%%%%%%%%%%%%%%%%%%%%%%%%%%%%%%%%%
\begin{figure*}[t]
\plotone{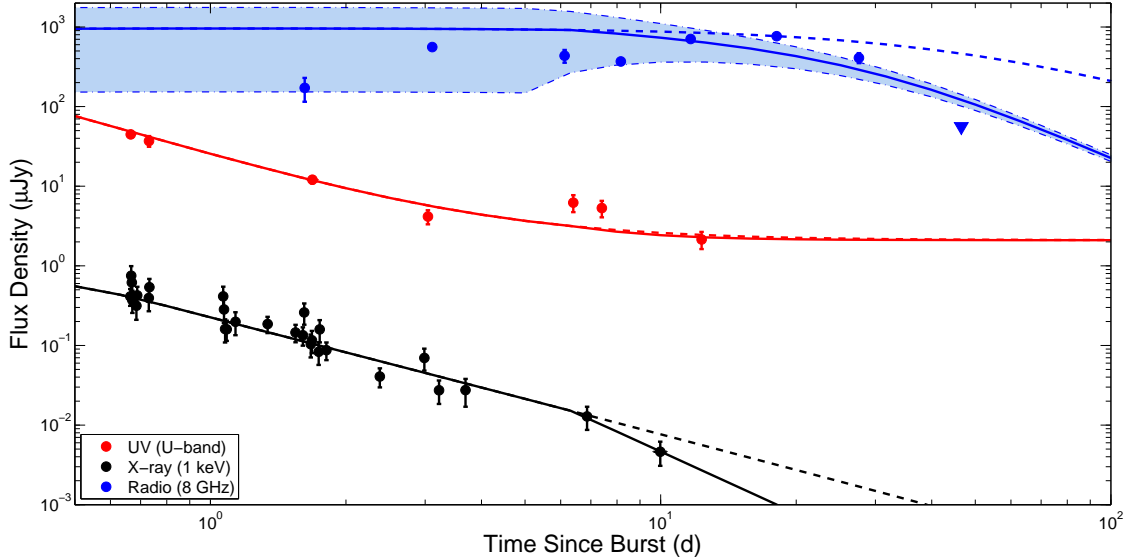}
%\plotone{../Figures/0328.eps}
\caption{The broadband radio (blue), UV (red), and X-ray (black) light
curve of \grbb.  The best-fit model is plotted in solid lines (see 
Table~\ref{tab:0328mod} for parameters).  The identical model
parameters for an isotropic explosion are plotted as the dashed lines.  
The radio light curve is not
very well fit at early times, although it likely suffers from strong
interstellar scintillation (light-blue shaded region).  In this case a jet break
 is 
required by the data to fall at $t \gtrsim 10$\,days.}
\label{fig:0328}
\end{figure*}
%%%%%%%%%%%%%%%%%%%%%%%%%%%%%%%%%%%%%%%%%%%%%%%%%%%%%%%%%%%%

%%%%%%%%%%%%%%%%%%%%%%%%%%%%%%%%%%%%%%%%%%%%%%%%%%%%%%%%%%%%%%%%%%%%%%%%%%
\subsubsection{Optical Spectroscopy}
\label{sec:0323spec}
We began spectroscopic observations of the optical afterglow of \grba\
with GMOS-S on 24 March 2009 ($\sim 29.93$\,hours after the GBM 
trigger; \citealt{GCN.9028}).  We first obtained 2 $\times$ 600\,s spectra
with the B600 grating and a central wavelength of 6000\,\AA, providing
coverage over the range $\sim 4500$--7500\,\AA\ with a resolution of 2.5\,\AA.  
Immediately following these exposures, we obtained 2 $\times$ 600\,s 
spectra with the R400 grating and a central wavelength of 8000\,\AA, 
providing coverage of $\sim 6000$--10000\,\AA\ with a resolution of 
5.0\,\AA.

The spectra were reduced in a standard manner using routines from the IRAF
\texttt{gemini} and \texttt{specred} packages (see, e.g., \citealt{cfp+08}
and references therein for details).  Wavelength calibration was performed
relative to CuAr lamps and then adjusted based on measured night-sky emission 
lines.  The resulting root-mean square wavelength uncertainty was 
$\lesssim 0.25$\,\AA\ for the spectra with the B600 grating and $\lesssim 
0.40$\,\AA\ for the spectra taken with the R400 grating.  Telluric features 
were removed using the continuum from well-exposed spectrophotometric 
standard stars (e.g., \citealt{wh88,mfh+00}).

Flux calibration was performed relative to the spectrophotometric standard
star LTT\,7379 \citep{sb83,bs84}.  We caution, however, that the standard-star 
observations were conducted on different nights from the GRB
observations (31 August 2009 for the B600 grating; 4 August 2009 for the
R400 grating), so the absolute flux calibration is somewhat uncertain 
(estimated to be $\sim 30$\%).

The resulting spectrum of the afterglow of \grba\ is shown in 
Figure~\ref{fig:0323spec}.  The broad absorption feature at $\lambda \approx 
5570$\,\AA\ is produced by Ly$\alpha$ at $z \approx 3.6$.  The 
spectrum blueward of this wavelength is dominated by absorption from
the Ly$\alpha$ forest, while the lack of Ly$\alpha$ forest features redward
of this transition indicates that it corresponds to the redshift of the GRB host
galaxy.  Unfortunately, the GMOS CCD chip gap over the range
5500--5525\,\AA\ precludes an accurate measurement of the Ly$\alpha$
profile and hence a determination of the \ion{H}{1} column density.

Redward of Ly$\alpha$, we find a series of strong (rest-frame equivalent
width $W_{0} \ge 1$\,\AA) absorption features superposed on a relatively
flat, featureless continuum.  In particular, we identify \ion{Si}{2} 
$\lambda$1260, \ion{Si}{2}$^{*}$ $\lambda$1264, \ion{O}{1} $\lambda$1302, 
\ion{Si}{2} $\lambda$1304, \ion{Si}{2}$^{*}$ $\lambda$1309, 
\ion{C}{2} $\lambda$1334, \ion{C}{2}$^{*}$ $\lambda$1335,
\ion{Si}{4} $\lambda\lambda$1393, 1402, \ion{Si}{2} $\lambda$1526,
\ion{Si}{2}$^{*}$ $\lambda$1533, and \ion{C}{4} $\lambda\lambda$1548, 1550
at $z = 3.568 \pm 0.004$.  Excited fine structure lines are indicative
of UV pumping and further suggest these features arise in the host
galaxy of \grba.  The strongest features 
exhibit a complex velocity structure that is only marginally resolved at the
resolution of our B600 spectra.  We also detect intervening absorption 
systems of  \ion{C}{4} $\lambda\lambda$1548, 1550 at $z = 3.379$ and 
\ion{Mg}{2} $\lambda\lambda$2796, 2803 at $z = 2.101$.  A more
thorough analysis of the host-galaxy properties of \grba\ will be presented
in a subsequent work.  

At $z = 3.568$, the isotropic prompt energy release from
\grba\ in the rest-frame 1--10$^{4}$\,keV bandpass is
$E_{\gamma,\mathrm{iso}} = (3.99 \pm 0.53) \times 10^{54}$\,erg.
Using the formulation of \citet{ls01}, the lower limit on the outflow
Lorentz factor (assuming a nonthermal spectrum up to
$E_{\mathrm{obs}} \approx 500$\,keV) is $\Gamma_{0} \gtrsim 600$.

%%%%%%%%%%%%%%%%%%%%%%%%%%%%%%%%%%%%%%%%%%%%%%%%%%%%%%%%%%%%%%%%%%%%%%%%%
\subsection{\grbb}
\label{sec:0328}
%%%%%%%%%%%%%%%%%%%%%%%%%%%%%%%%%%%%%%%%%%%%%%%%%%%%%%%%%%%%%%%%%%%%%%%%%

%%%%%%%%%%%%%%%%%%%%%%%%%%%%%%%%%%%%%%%%%%%%%%%%%%%%%%%%%%%%%%%%%%%%%%%%%
\subsubsection{High-Energy Properties}
\label{sec:0328high}
\grbb\ triggered the \fermi\ GBM at 09:36:46 on 28 March 2009
\citep{GCN.9044}.  In the GBM bandpass, the light curve is multipeaked
with a duration of $t_{90}\approx 70$\,s.  \grbb\ was also detected at 
MeV energies by several IPN satellites \citep{GCN.9049}, including
Konus-\textit{Wind} \citep{GCN.9050}.

A spectrum consisting of the first $\sim 30$\,s of GBM data
(containing the brightest part of the high-energy light curve) is well fit
by a Band function with $\alpha = -0.93 \pm 0.02$, $\beta = -2.2 \pm 0.1$,
$E_{\mathrm{p}} = 653 \pm 45$\,keV \citep{GCN.9057}.  The resulting
fluence in the 8\,keV to 40\,MeV (observer frame) bandpass is 
$f_{\gamma} = (9.5 \pm 1.0) \times 10^{-5}$\,erg\,cm$^{-2}$.  These
results are all consistent with the analogous values derived by the
Konus-\textit{Wind} instrument, which include data out to $t_{0} +
73$\,s.  Given the wider bandpass of the GBM and the higher precision
of its fluence measurement, we adopt this value for remainder of this
work.  Using the observed redshift of 0.736 ($\S$~\ref{sec:0323spec}),
we find a prompt fluence of $f_{\gamma} = (7.3 \pm 0.8) \times
10^{-5}$\,erg\,cm$^{-2}$ in the 1--10$^{4}$\,keV rest-frame bandpass.

\grbb\ was also detected at GeV energies by the \fermi-LAT.  Much
like \grba, flux in the LAT band from \grbb\ is detected out to 
$\sim 900$\,s after the GBM trigger \citep{GCN.9077}.  Many photons
with energies above 1\,GeV are detected from the direction of \grbb;
however, they are all detected well after the end of the MeV
emission \citep{GCN.9077,p09}.  The highest energy photon detected
from the direction of \grbb\ was measured at $\sim 5$\,GeV
($t_{0} + 798$\,s), while the most energetic photon detected during the 
prompt emission had $E \approx 700$\,keV at $t_{0} + 60$\,s \citep{p09}.

%%%%%%%%%%%%%%%%%%%%%%%%%%%%%%%%%%%%%%%%%%%%%%%%%%%%%%%%%%%
\begin{figure*}[t]
\plotone{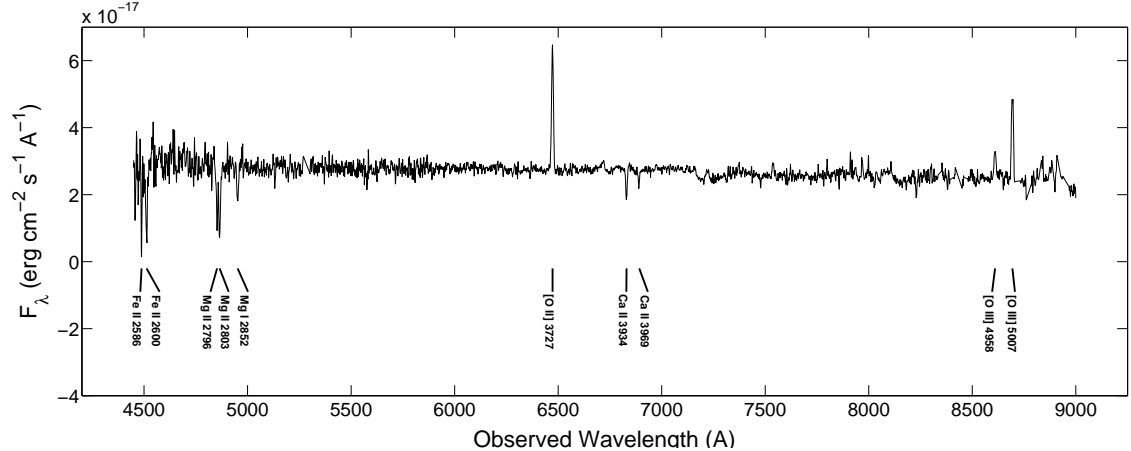}
%\plotone{../Figures/0328_spec.eps}
\caption{GMOS-S optical spectrum of the afterglow of \grbb.  We identify a series
of forbidden O emission lines and Fe, Mg, and Ca absorption features at a common
redshift of $0.7357 \pm 0.0004$.  We have only plotted the
wavelength range 4500--9000\,\AA, due to the
lower S/N in the blue (spectrograph throughput) and
the red (fringing).}
\label{fig:0328spec}
\end{figure*}
%%%%%%%%%%%%%%%%%%%%%%%%%%%%%%%%%%%%%%%%%%%%%%%%%%%%%%%%%%%

%%%%%%%%%%%%%%%%%%%%%%%%%%%%%%%%%%%%%%%%%%%%%%%%%%%%%%%%%%%%%%%%%%%%%
\subsubsection{Afterglow Observations}
\label{sec:0328ag}
The \swift\ XRT began target-of-opportunity observations of \grbb\
at 01:26 on 29 March 2009 ($\sim 15.9$\,hours after the GBM trigger).
A fading X-ray source at $\alpha = 6^{\mathrm{h}} 02^{\mathrm{m}}
39.^{\mathrm{s}}58$, $\delta = -41^{\circ} 52\arcmin 57\farcs5$ (J2000.0; 
$2\farcs0$ containment radius) was promptly identified as the
X-ray afterglow \citep{GCN.9045,GCN.9046}.  The resulting X-ray
light curve, showing the afterglow evolution over the following 10 days, 
is plotted in Figure~\ref{fig:0328}.

The Ultraviolet Optical Telescope (UVOT; \citealt{rkm+05}) onboard \swift\
began settled observations of the field of \grbb\ approximately 16\,hours after
the \fermi\ trigger.  The optical afterglow, detected in both the $U$-band
and white filters, was identified shortly thereafter \citep{GCN.9048}.  
UVOT observations of the afterglow continued for $\sim 2$ weeks, almost
exclusively in these two filters.  We have downloaded the UVOT $U$-band data
from the HEASARC archive\footnote{See \texttt{http://heasarc.gsfc.nasa.gov}.} 
and conducted photometry with the images following the technique described by 
\citet{ljf+06}.  Given the large uncertainty associated with flux calibration 
for the white filter, we have not included data taken in this band.  The 
results of this analysis, not including a correction for foreground Galactic 
extinction ($E[B-V] = 0.057$\,mag; \citealt{sfd98}), are shown in 
Table~\ref{tab:0328opt}.

We obtained a single $i^{\prime}$ image of the afterglow of
\grbb\ with GMOS-S on 29 April 2009.  This image was reduced in an 
identical manner to our observations of \grba, and the zero point was 
calculated using predetermined values from the Gemini 
website\footnote{\texttt{http://www.gemini.edu/sciops/instruments/gmos/?q=sciops/\\instruments/gmos}.}.
We have also included the simultaneous GROND optical and NIR 
measurements from \citet{GCN.9054} in our modeling, and therefore display 
them in Table~\ref{tab:0328opt} as well.

Finally, we began observing the afterglow of \grbb\ with the VLA on 
30 March 2009 \citep{GCN.9060} and continued for nearly two months.  
The data were reduced in a manner identical to that described in 
\S~\ref{sec:0323ag}.  The results of our radio campaign are displayed
in Table~\ref{tab:0328rad} (individual epochs) and
plotted in Figure~\ref{fig:0328} (combined epochs). 

%%%%%%%%%%%%%%%%%%%%%%%%%%%%%%%%%%%%%%%%%%%%%%%%%%%%%%%%%%%%
\begin{deluxetable}{l r c c}
\tablewidth{0pt}
\tablecaption{VLA Radio Observations of GRB\,090328}
\tablehead{
\colhead{Date} & \colhead{$\Delta{t}$} &\colhead{$\nu$} &\colhead{$f_\nu$}\\
\colhead{(UT)} & (days) & (GHz) & ($\mu$Jy)
}
\startdata
Mar 30.02 & 1.62 & 8.46 & 172 $\pm$ 57 \\
Mar 30.99 & 2.59 & 8.46 & 337 $\pm$ 60 \\
Apr 1.02 & 3.62 & 8.46 & 783 $\pm$ 57 \\
Apr 3.03 & 5.63 & 8.46 & 195 $\pm$ 34 \\
Apr 4.01 & 6.61 & 8.46 & 674 $\pm$ 155 \\
Apr 5.09 & 7.69 & 8.46 & 214 $\pm$ 41 \\
Apr 6.02 & 8.62 & 8.46 & 523 $\pm$ 57 \\
Apr 6.97 & 9.57 & 8.46 & 809 $\pm$ 39 \\
Apr 11.14 & 13.74 & 8.46 & 603 $\pm$ 67 \\
Apr 14.00 & 16.60 & 8.46 & 643 $\pm$ 59 \\
Apr 16.99 & 19.59 & 8.46 & 886 $\pm$ 58 \\
Apr 24.97 & 27.57 & 8.46 & 410 $\pm$ 59 \\
May 9.96 & 42.56 & 8.46 & 78 $\pm$ 43 \\
May 11.86 & 44.46 & 8.46 & 21 $\pm$ 64 \\
May 17.92 & 50.52 & 8.46 & -20 $\pm$ 40 \\
\enddata
\label{tab:0328rad}
\end{deluxetable}
%%%%%%%%%%%%%%%%%%%%%%%%%%%%%%%%%%%%%%%%%%%%%%%%%%%%%%%%%%%

%%%%%%%%%%%%%%%%%%%%%%%%%%%%%%%%%%%%%%%%%%%%%%%%%%%%%%%%%%%%%%%%%%%%%%%%%%
\subsubsection{Optical Spectroscopy}
\label{sec:0328spec}
We began spectroscopic observations of \grbb\ with GMOS-S at 00:05 on 30 April
2009 ($\sim 38.5$\,hours after the GBM trigger; \citealt{GCN.9053}).  We 
obtained 2 $\times$ 1500\,s spectra with the R400 grating, the first 
with a central wavelength of 6000\,\AA\ (providing coverage of 
$\sim 4000$--8000\AA), and the second with a central wavelength of 
8000\,\AA\ (providing coverage of $\sim 6000$--10000\,\AA).  The 
spectra were reduced in the manner described in 
\S~\ref{sec:0323spec}.  Flux calibration was performed relative to 
the standard star LTT\,7379 taken with identical instrumental setups on 
4 August 2009.

The resulting spectrum of \grbb\ is shown in Figure~\ref{fig:0328spec}.  
Superposed on a relatively flat continuum, we identify strong emission lines 
of oxygen ([\ion{O}{2}] $\lambda$3727, [\ion{O}{3}] $\lambda\lambda$4959, 
5007) as well as a series of narrow absorption features (\ion{Fe}{2}
$\lambda$2586, \ion{Fe}{2} $\lambda$2600, \ion{Mg}{2} $\lambda\lambda$2796, 
2803, \ion{Mg}{1} $\lambda$2852, and \ion{Ca}{2} $\lambda\lambda$ 3934, 3969), 
all at $z = 0.7357 \pm 0.0004$.

At this redshift, the prompt isotropic gamma-ray energy release from
\grbb\ in the 1--10$^{4}$\,keV bandpass is $E_{\gamma,\mathrm{iso}} =
(1.34 \pm 0.14) \times 10^{53}$\,erg.  Using the formulation of
\citet{ls01}, the lower limit on the outflow Lorentz factor (assuming
a nonthermal spectrum up to $E_{\mathrm{obs}} \approx 700$\,keV) is 
$\Gamma_{0} \gtrsim 200$.

%%%%%%%%%%%%%%%%%%%%%%%%%%%%%%%%%%%%%%%%%%%%%%%%%%%%%%%%%%%%%%%%%%%%%%%%%%%
\subsection{\grbc}
\label{sec:0902b}
%%%%%%%%%%%%%%%%%%%%%%%%%%%%%%%%%%%%%%%%%%%%%%%%%%%%%%%%%%%%%%%%%%%%%%%%%%%

%%%%%%%%%%%%%%%%%%%%%%%%%%%%%%%%%%%%%%%%%%%%%%%%%%%%%%%%%%%
\begin{figure*}[t]
\plotone{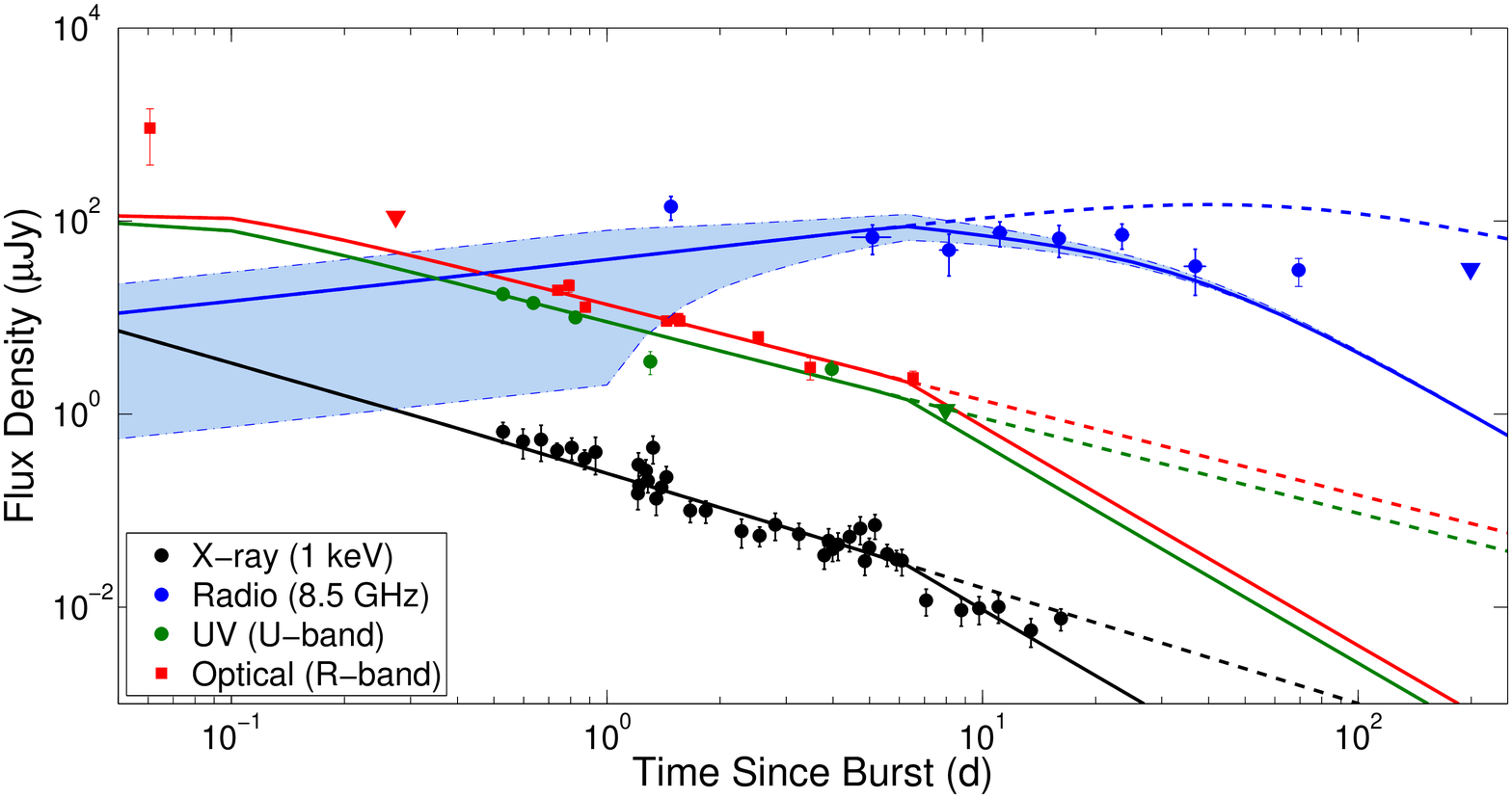}
%\plotone{../Figures/0902.eps}
\caption{The broadband radio (blue), optical (red), UV (cyan), and X-ray 
(black) light curves of \grbc.  The best-fit model is plotted in solid lines 
(see Table~\ref{tab:0902mod} for parameters).  The identical model
parameters for an isotropic explosion are plotted as the dashed lines.  
The strength of the possible
modulation of the radio afterglow caused by interstellar scintillation
(e.g., \citealt{fwk00}) is indicated by the light-blue shaded region.  
As suggested by \citet{psp+10}, the early $(t \lesssim 0.3$\,day)
optical (and, to a lesser extent, radio) data are significantly larger
than predicted by our forward-shock afterglow models.
The most likely explanation is the presence of reverse-shock emission.  
The lack of a bright radio afterglow at late times suggests the 
presence of a jet break at $t \approx 6$\,days.}
\label{fig:0902}
\end{figure*}
%%%%%%%%%%%%%%%%%%%%%%%%%%%%%%%%%%%%%%%%%%%%%%%%%%%%%%%%%%%%

%%%%%%%%%%%%%%%%%%%%%%%%%%%%%%%%%%%%%%%%%%%%%%%%%%%%%%%%%%%%%%%%%%%%%%%%%%%
\subsubsection{High-Energy Properties}
\label{sec:0902bhigh}
At 11:05:08.31 on 2 September 2009, the \fermi-GBM triggered and located
\grbc\ \citep{GCN.9866}.  In the GBM bandpass, the light curve consisted
of a bright, multipeaked pulse with a duration $t_{90} \approx 21$\,s.  
\grbc\ was also detected in a similar bandpass by \textit{Suzaku}-WAM 
\citep{GCN.9897}.

Furthermore, \grbc\ was bright enough to be detected at GeV energies by
the \fermi-LAT \citep{GCN.9867}.  Like the other events in our sample,
LAT emission was seen out to $t_{0} + 1000$\,s, at late times decaying
like a power law with temporal index $\alpha_{\mathrm{LAT}} \approx 1.5$
\citep{aaa+09b}.  \grbc\ is responsible for the highest energy photon
detected from a GRB to date, with $E = 33.4_{-3.5}^{+3.7}$\,GeV at
$t_{0} + 82$\,s.  Within the MeV prompt emission phase, the highest energy
photon was measured at $E = 11.16_{-0.58}^{+1.48}$\,GeV \citep{aaa+09b}.

Unlike the other events considered here, the prompt high-energy spectrum of 
\grbc\ is not adequately described by a single Band function.  While the 
100--1000\,keV bandpass is reasonably well fit by such a model, 
the spectrum exhibits excess emission at both low ($\lesssim 100$\,keV)
and high ($\gtrsim 10$\,MeV) energies.  \citet{aaa+09b} have suggested that
the high-energy spectrum of \grbc\ can be reproduced as the sum of two
components: a Band function peaking at $\sim 700$\,keV, and a single
power-law (photon index $\Gamma \equiv \beta + 1 = 1.93$) extending over 
the entire GBM+LAT bandpass.  In this model, the power-law component
accounts for $\sim 24$\% of the total 10\,keV to 10\,GeV fluence.  The
physical mechanism responsible for this complex spectrum is still not 
entirely understood; possible explanations include a hadronic origin
[either proton synchrotron radiation \citep{rdf09} or photohadronic 
interactions \citep{agm09}] or thermal emission from the jet photosphere
\citep{mr00,r04,raz+10}.

Using their two-component spectrum, \citet{aaa+09b} report a total fluence
of $f_{\gamma} = (4.59 \pm 0.05) \times 10^{-4}$\,erg\,cm$^{-2}$ in the 
8\,keV to 30\,GeV (observer frame) bandpass.  At a redshift of 1.8229
(\S~\ref{sec:0902bspec}), this corresponds to a prompt fluence of 
$(3.83 \pm 0.05) \times 10^{-4}$\,erg\,cm$^{-2}$ in the 
1--10$^{4}$\,keV rest-frame bandpass.

%%%%%%%%%%%%%%%%%%%%%%%%%%%%%%%%%%%%%%%%%%%%%%%%%%%%%%%%%%%%%%%%%%%%%%%%%%%
\subsubsection{Afterglow Observations}
\label{sec:0902bag}
The \swift\ XRT began target-of-opportunity observations of the field of
\grbc\ at 23:36 on 2 September 2009 ($\sim 12.5$\,hr after the GBM
trigger).  A fading X-ray source at $\alpha = 17^{\mathrm{h}} 39^{\mathrm{m}}
45.^{\mathrm{s}}26$, $\delta = +27^{\circ} 19\arcmin 28\farcs1$ (J2000.0; 
$2\farcs1$ containment radius) was promptly identified as the X-ray
afterglow \citep{GCN.9868,GCN.9871,GCN.9876}.  The resulting X-ray
light curve is shown in Figure~\ref{fig:0902}.

The \swift\ UVOT began concurrently observing the field of \grbc\ and
first reported the detection of a candidate optical afterglow consistent
with the X-ray position \citep{GCN.9869}.  Subsequent observations
revealed that the candidate had faded, confirming its association with
\grbc\ \citep{GCN.9877}.  In Table~\ref{tab:0902opt} we present UVOT 
$U$-band observations of \grbc, reduced in an identical manner to those
described in \S~\ref{sec:0328ag}.

We obtained a single $r^{\prime}$ image of the afterglow of
\grbc\ with GMOS-N on 3 September 2009.  This
image was reduced in the same manner as those of the other
events, and the resulting photometry is presented in 
Table~\ref{tab:0902opt}.  We have included optical and NIR photometry
of \grbc\ from \citet{psp+10} in our modeling, and therefore display them 
in Table~\ref{tab:0902opt} as well.

Finally, we began observing the afterglow of \grbc\ with the VLA on 3 
September 2009 \citep{GCN.9889} and continued for $\sim 5$ months.  
The data were reduced as described in 
$\S$~\ref{sec:0328ag}.  The results of our radio campaign are displayed
in Tables~\ref{tab:0902rad} (individual epochs) and plotted in 
Figure~\ref{fig:0902} (combined epochs). We have also
included the early radio detection at 4.8\,GHz reported by the WSRT 
\citep{GCN.9883} in Table~\ref{tab:0902rad}.

%%%%%%%%%%%%%%%%%%%%%%%%%%%%%%%%%%%%%%%%%%%%%%%%%%%%%%%%%%%
\begin{figure*}
\plotone{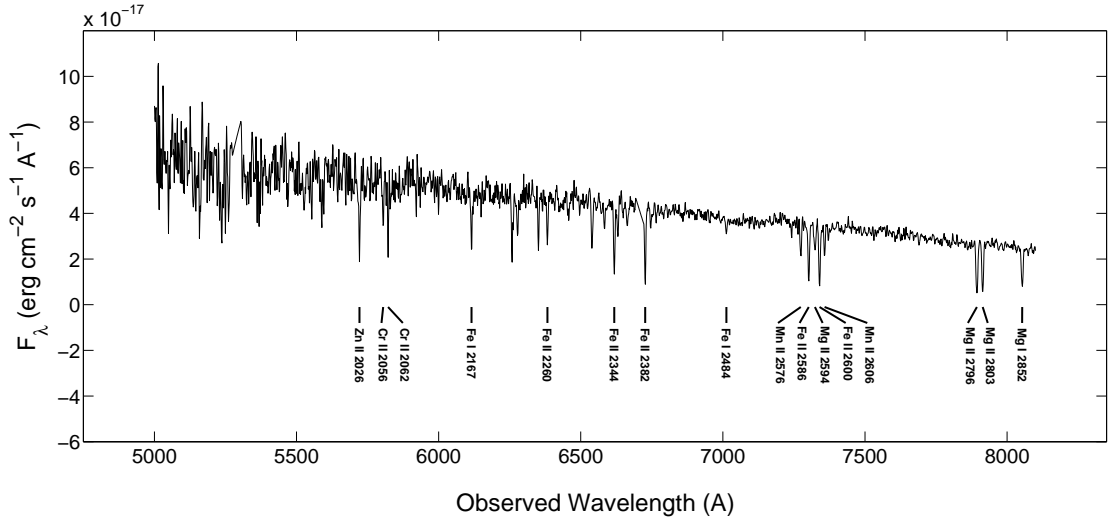}
%\plotone{../Figures/0902b_spec.eps}
\caption{GMOS-N optical spectrum of the afterglow of \grbc.  We identify a series
of absorption features from neutral and singly ionized Mg, Mn, Fe, and Cr
at a common redshift of $1.8229 \pm 0.0004$.  We have not plotted the
shortest wavelengths due to the decreased spectrograph throughput.}
\label{fig:0902bspec}
\end{figure*}
%%%%%%%%%%%%%%%%%%%%%%%%%%%%%%%%%%%%%%%%%%%%%%%%%%%%%%%%%%%

%%%%%%%%%%%%%%%%%%%%%%%%%%%%%%%%%%%%%%%%%%%%%%%%%%%%%%%%%%%%%%
\begin{deluxetable}{lrccc}
  \tablewidth{0pt}
  \tablecaption{Radio Observations of \grbc}
  \tablehead{
    \colhead{Date} & \colhead{$\Delta{t}$} &\colhead{$\nu$} &
    \colhead{$f_{\nu}$} & \colhead{Facility} \\ 
    \colhead{(UT)} & (days) & (GHz) & ($\mu$Jy)
  }
  \startdata
  2009 Sep 3.77 & 1.31 & 4.8 & $111 \pm 28$ & WSRT \tablenotemark{a} \\
  2009 Sep 3.94 & 1.48 & 8.46 & $141 \pm 39$ & VLA \\
  2009 Sep 7.05 & 4.59 & 8.46 & $13 \pm 31$ & VLA \\
  2009 Sep 8.05 & 5.59 & 8.46 & $130 \pm 34$ & VLA \\
  2009 Sep 10.15 & 7.69 & 8.46 & $10 \pm 32$ & VLA \\
  2009 Sep 11.05 & 8.59 & 8.46 & $80 \pm 32$ & VLA \\
  2009 Sep 13.10 & 10.64 & 8.46 & $99 \pm 31$ & VLA \\
  2009 Sep 14.14 & 11.60 & 8.46 & $71 \pm 33$ & VLA \\
  2009 Sep 18.04 & 15.50 & 8.46 & $52 \pm 32$ & VLA \\
  2009 Sep 19.00 & 16.46 & 8.46 & $89 \pm 36$ & VLA \\
  2009 Sep 25.09 & 22.51 & 8.46 & $26 \pm 29$ & VLA \\
  2009 Sep 27.08 & 24.50 & 8.46 & $67 \pm 29$ & VLA \\
  2009 Oct 7.01 & 34.43 & 8.46 & $38 \pm 28$ & VLA \\
  2009 Oct 9.01 & 36.43 & 8.46 & $66 \pm 27$ & VLA \\
  2009 Oct 11.81 & 39.23 & 8.46 & $21 \pm 31$ & VLA \\
  2009 Nov 6.90 & 65.44 & 8.46 & $9 \pm 20$ & VLA \\
  2009 Nov 7.77 & 66.31 & 8.46 & $22 \pm 19$ & VLA \\
  2009 Nov 9.00 & 67.54 & 8.46 & $48 \pm 19$ & VLA \\
  2009 Nov 14.88 & 73.42 & 8.46 & $31 \pm 21$ & VLA \\
  2010 Mar 20.62 & 199.16 & 8.46 & $18 \pm 16$ & VLA \\
  \enddata
\tablenotetext{a}{Reference: \citet{GCN.9883}.}
\label{tab:0902rad}
\end{deluxetable}
%%%%%%%%%%%%%%%%%%%%%%%%%%%%%%%%%%%%%%%%%%%%%%%%%%%%%%%%%%%%%%%

%%%%%%%%%%%%%%%%%%%%%%%%%%%%%%%%%%%%%%%%%%%%%%%%%%%%%%%%%%%%%%%%%%%%%%%%%%%
\subsubsection{Optical Spectroscopy}
\label{sec:0902bspec}
We began spectroscopic observations of \grbc\ with GMOS-N at 06:29 on 3 
September 2009 ($\sim 19.4$\,hours after the GBM trigger; \citealt{GCN.9873}).
We obtained 2 $\times$ 900\,s exposures, both with the R400 grating and a
central wavelength of 6000\,\AA, providing coverage of $\sim 
4000$--8000\,\AA.
The spectra were reduced as described in 
\S~\ref{sec:0323spec} and \S~\ref{sec:0328spec}.  Flux calibration was
performed relative to spectra of the standard star Feige\,34 \citep{o90}
taken with the same instrumental setup as on 1 May 2008.

The resulting spectrum of \grbc\ is shown in Figure~\ref{fig:0902bspec}.
Super-imposed on a smooth power-law continuum, we identify strong 
absorption features corresponding to \ion{Mg}{1} $\lambda$2852, 
\ion{Mg}{2} $\lambda\lambda$2796, 2803, \ion{Mn}{2} $\lambda$2606, 
\ion{Fe}{2} $\lambda$2600, \ion{Mn}{2} $\lambda$2594, 
\ion{Fe}{2} $\lambda$2586, \ion{Mn}{2} $\lambda$2576, \ion{Fe}{1}
$\lambda$2484, \ion{Fe}{2} $\lambda$2382, \ion{Fe}{2} $\lambda$2344, 
\ion{Fe}{2} $\lambda$2260, \ion{Fe}{1} $\lambda$2167,
\ion{Cr}{2} $\lambda\lambda$2056, 2062, and \ion{Mg}{1} $\lambda$2026, 
all at a common redshift of $1.8229 \pm 0.0004$.  

The lack of evidence for Ly$\alpha$ absorption down to $\lambda 
\lesssim 4000$\,\AA\ implies an upper bound on the redshift of the
GRB host galaxy of $z \lesssim 2.3$.  Together with the strength of 
the above features (in particular \ion{Fe}{1} $\lambda$2167), we 
consider it quite likely that the observed
system at $z = 1.8229$ derives from the GRB host galaxy.

At this redshift, the prompt isotropic gamma-ray energy release in
the 1--10$^{4}$\,keV rest-frame bandpass is $E_{\gamma,\mathrm{iso}} = 
(3.20 \pm 0.04) \times 10^{54}$\,erg.  Assuming a nonthermal 
spectrum up to $E_{\mathrm{obs}} = 11$\,GeV, \citet{aaa+09b} have
determined a lower limit to the initial outflow Lorentz factor of
$\Gamma_{0} \gtrsim 1000$.

%%%%%%%%%%%%%%%%%%%%%%%%%%%%%%%%%%%%%%%%%%%%%%%%%%%%%%%%%%%%%%%%%%%%%%%%%%
\subsection{\grbd}
\label{sec:0926a}
%%%%%%%%%%%%%%%%%%%%%%%%%%%%%%%%%%%%%%%%%%%%%%%%%%%%%%%%%%%%%%%%%%%%%%%%%%

%%%%%%%%%%%%%%%%%%%%%%%%%%%%%%%%%%%%%%%%%%%%%%%%%%%%%%%%%%%%%%%%%%%%%%%%%%
\subsubsection{High-Energy Properties}
\label{sec:0926ahigh}
\grbd\ triggered the GBM on \fermi\ at 04:20:26.99 on 26 September
2009 \citep{GCN.9933}.  The light curve consists of a single pulse
with duration $t_{90} \approx 20$\,s in the GBM bandpass.  The prompt 
emission was sufficiently bright to trigger several additional satellite 
instruments, including \textit{Suzaku}-WAM \citep{GCN.9951}, 
Konus-\textit{Wind} \citep{GCN.9959}, and RT-2 on CORONAS-PHOTON
\citep{GCN.10009}.  

Simultaneously fitting both the GBM and LAT (see below) data from 
$t_{0}$ to $t_{0} + 21$\,s, \citet{GCN.9972} report that the spectrum is
reasonably well modeled by a Band function with $\alpha = -0.693 \pm 0.009$,
$\beta = -2.342 \pm 0.011$, and $E_{\mathrm{p}} = 268 \pm 4$\,keV.
The corresponding 10\,keV to 10\,GeV fluence (observer frame) is
$f_{\gamma} = (2.47 \pm 0.03) \times 10^{-4}$\,erg\,cm$^{-2}$.  These
results differ only slightly from the values reported from other
satellites.  At $z = 2.1062$ \citep{GCN.9942}, the corresponding
prompt fluence in the rest-frame 1--10$^{4}$\,keV bandpass is 
$(1.73 \pm 0.03) \times 10^{-4}$\,erg\,cm$^{-2}$.

\grbd\ was also detected by the \fermi-LAT.  The emission in the 
LAT bandpass lasted at least 400\,s, with possible indications
of significant flux out to $\sim 1000$\,s after the GBM trigger.
Many photons with $E > 1$\,GeV were detected from the position of
\grbd, with the highest energy measured of 20\,GeV at $t_{0} + 26$\,s
\citep{GCN.9934}.

%%%%%%%%%%%%%%%%%%%%%%%%%%%%%%%%%%%%%%%%%%%%%%%%%%%%%%%%%%%%%%%%%%%%%%%%%%
\subsubsection{Afterglow Observations}
\label{sec:0926aag}
\begin{figure}[t]
\epsscale{1.2}
\plotone{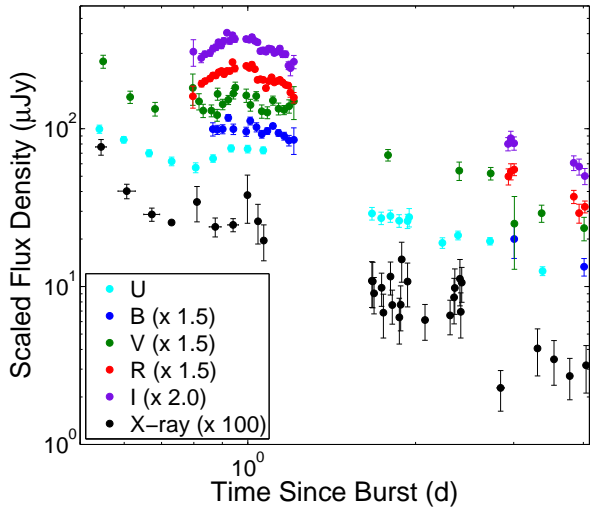}
%\plotone{../Figures/0926_opt.eps}
\caption{Early optical and X-ray afterglow of \grbd.  After an initial
decline in the $U$ and $V$ filters, the afterglow experiences a prominent 
rebrightening in all filters, peaking at $t \approx 1$\,day.  Such behavior is 
difficult to reconcile with standard afterglow theory, and may suggest
a late-time ($t \gg \Delta t_{\mathrm{GRB}}$)
injection of energy from the central engine \citep{rm98}.}
\label{fig:0926opt}
\end{figure}

\begin{figure}[t]
\epsscale{1.2}
\plotone{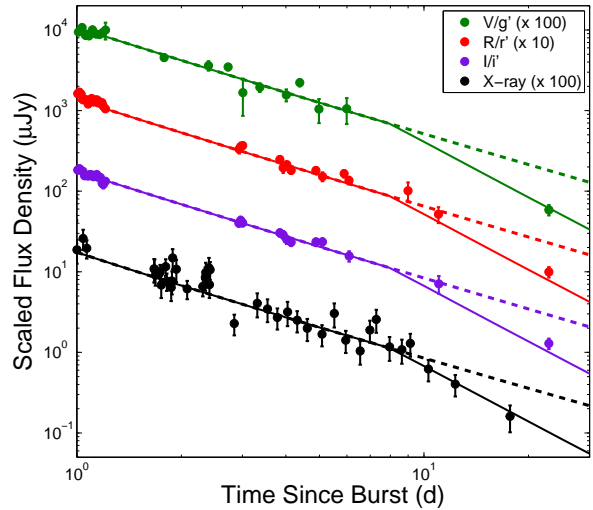}
%\plotone{../Figures/0926_late.eps}
\caption{Late-time afterglow and model of \grbd.  Both the X-ray and optical 
($g^{\prime} r^{\prime} i^{\prime}$) bandpasses exhibit a steepening decline
at $t \approx 10$\,days, strongly indicative of a jet break.}
\label{fig:0926late}
\end{figure}

The \swift\ XRT began observing the field of \grbd\ at 17:17 on 26
September 2009 ($\sim 13$\,hours after the GBM trigger).  A fading
X-ray counterpart at $\alpha = 23^{\mathrm{h}} 33^{\mathrm{m}} 36.^{\mathrm{s}}18$,
$\delta = -66^{\circ} 19\arcmin 25\farcs9$ (J2000.0, $1\farcs5$ 
containment radius) was promptly identified in the XRT data
\citep{GCN.9936,GCN.9961}.  The XRT observed \grbd\ for the next 3 weeks,
and we plot the X-ray light curve in Figures~\ref{fig:0926opt} and \ref{fig:0926late}.

%%%%%%%%%%%%%%%%%%%%%%%%%%%%%%%%%%%%%%%%%%%%%%%%%%%%%%%%%
%%%%%%%%%%%%%%%%%%%%%%%%%%%%%%%%%%%%%%%%%%%%%%%%%%%%%%%%%

Under the control of Skynet, four of the 16\,inch diameter PROMPT 
telescopes \citep{rnm+05} at Cerro Telolo Inter-American Observatory
(CTIO) observed the \fermi-LAT localization of \grbd\ beginning 19.0 hours 
after the GBM trigger in the $B$-, $V$-, $R$-, and $I$-band filters.  
Within the {\it Swift}-XRT localization, we identified an uncatalogued and 
fading source as the optical afterglow of \grbd\
\citep{GCN.9937,GCN.9953}.  Individual images were automatically 
reduced using our custom, IRAF-based reduction pipeline and then
astrometrically aligned and stacked.  We subsequently measured the afterglow 
flux with aperture photometry, where the inclusion radius was 
approximately matched to the FWHM of the PSF.
PROMPT continued to observe the field for ten more nights 
\citep{GCN.9982,GCN.9984,GCN.10003}.

We also obtained 3 epochs of $I$- and $J$-band imaging of the afterglow
of \grbd\ using the ANDICAM (A Novel Dual Imaging CAMera) instrument mounted
on the 1.3\,m telescope at 
CTIO\footnote{\texttt{http://www.astronomy.ohio-state.edu/ANDICAM}.}.  
This telescope is operated as part of the Small and Moderate Aperture 
Research Telescope System (SMARTS) 
consortium\footnote{\texttt{http://www.astro.yale.edu/smarts}.}.
Each epoch consisted of 6 individual 360 s $I$-band observations 
and 30 individual 60 s $J$-band observations.  Between optical exposures, 
the telescope was slightly offset and the individual $J$-band exposures 
were additionally dithered via an internal tilting mirror system.  
Standard data reduction was performed on these images, including cosmic ray 
rejection, overscan bias subtraction, zero subtraction, flat fielding
and sky subtraction to correct for the NIR background and the $I$-band
fringing.  For each epoch, the individual images were then aligned and 
averaged to produce a single frame in each band with summed exposure
times of 36 minutes in $I$ and 30 minutes in $J$.

Relative aperture photometry was performed on the SMARTS data, in
comparison with a number of nonvariable sources in the field of \grbd. 
The $I$-band field was photometrically calibrated by comparison (on 
photometric nights) with Landolt standard stars in the field of T 
Phe \citep{l92}.  $J$-band photometric calibration was performed 
using 2MASS \citep{scs+06} field stars.  

SMARTS $BVRI$ observations of the field of \grbd\ were also obtained on two 
photometric nights a few months after the GRB occurred (16 and 18 
December 2009).  For these observations, total summed exposure times 
amounted to 180 s in $BRI$ and 120\,s in $V$. The absolute photometry 
of the field was again established based on same-night observations
of the T Phe Landolt standard stars.  These observations were then
used to provide absolute calibration for the PROMPT observations of
\grbd.

We obtained additional late-time imaging of the field of \grbd\ on 
19 October 2009 with GMOS-S on Gemini South.  A total of 600\,s
of exposure time was obtained in the Sloan $g^{\prime}$-, $r^{\prime}$-,
and $i^{\prime}$-band filters.  The data were reduced in the manner
described in $\S$~\ref{sec:0323ag}, and calibrated in the same way
as the PROMPT and SMARTS observations.  

The results of our optical and NIR monitoring campaign of the afterglow
of \grbd, uncorrected for the modest amount of Galactic extinction 
[$E(B-V) = 0.024$\,mag; \citealt{sfd98}], are shown in 
Table~\ref{tab:0926aopt} and Figures~\ref{fig:0926opt} and 
\ref{fig:0926late}.

%%%%%%%%%%%%%%%%%%%%%%%%%%%%%%%%%%%%%%%%%%%%%%%%%%%%%%%%%
%%%%%%%%%%%%%%%%%%%%%%%%%%%%%%%%%%%%%%%%%%%%%%%%%%%%%%%%%

Finally, the field of \grbd\ was observed in the radio (5.5\,GHz)
on 1 October 2009 with the Australia Telescope Compact Array (ATCA).
No source was detected at the afterglow location to a 2$\sigma$
limit of $f_{\nu} < 1.5$\,mJy \citep{GCN.10020}.

%%%%%%%%%%%%%%%%%%%%%%%%%%%%%%%%%%%%%%%%%%%%%%%%%%%%%%%%%%%%%%%%%%%%%%%%%%
\subsubsection{Spectroscopy}
\label{sec:0926spec}
\citet{GCN.9942} obtained a spectrum of the afterglow of \grbd\ with the
X-Shooter instrument mounted on the 8\,m Very Large Telescope UT2.  Based
on the detection of a damped-Ly$\alpha$ system and many strong, narrow
absorption features, these authors derive a redshift of 2.1062 for
the host galaxy of \grbd.

At this redshift, the prompt isotropic gamma-ray energy release in
the 1--10$^{4}$\,keV bandpass is $E_{\gamma,\mathrm{iso}} = 
(1.89 \pm 0.03) \times 10^{54}$\,erg.  Assuming a nonthermal 
spectrum up to $E_{\mathrm{obs}} = 20$\,GeV, we infer a lower limit for
the initial Lorentz factor of $\Gamma_{0} \gtrsim 700$ \citep{ls01}.

%%%%%%%%%%%%%%%%%%%%%%%%%%%%%%%%%%%%%%%%%%%%%%%%%%%%%%%%%%%%%%%%%%%%%%%%%%
\section{Afterglow Modeling and Results}
\label{sec:results}
%%%%%%%%%%%%%%%%%%%%%%%%%%%%%%%%%%%%%%%%%%%%%%%%%%%%%%%%%%%%%%%%%%%%%%%%%%
In the standard ``fireball'' model of GRBs (e.g., \citealt{p05}), a compact
central engine (likely either a black hole or a rapidly spinning 
proto-neutron star) drives a collimated ($\theta \lesssim 10^{\circ}$), 
ultrarelativistic ($\Gamma_{0} \gtrsim 100$) outflow of matter and/or 
radiation.  Some dissipative process within the outflow (possibly collisionless
shocks) gives rise to the prompt gamma-ray emission, converting some 
fraction ($\eta_{\gamma} \equiv E_{\gamma,\mathrm{iso}} / (E_{\gamma,\mathrm{iso}} +
E_{\mathrm{KE,iso}}$) of the total relativistic energy to high-energy radiation.

The outflow is ultimately slowed as it sweeps up and shock heats the 
circumburst medium, and synchrotron radiation from electrons accelerated
at the shock front results in the broadband afterglow.  The resulting
spectrum is well described as a series of broken power laws with three
characteristic frequencies: $\nu_{a}$, the frequency below which the radiation
is self-absorbed; $\nu_{m}$, the characteristic frequency of the emitting
electrons; and $\nu_{c}$, the frequency above which electrons are able to
cool efficiently through radiation (e.g., \citealt{gs02}).

The temporal evolution of the afterglow depends on the density profile
of the circumburst medium.  We consider here two possibilities: a 
constant-density circumburst medium [$\rho(r) \propto r^{0}$], as would be 
expected in an environment similar to the interstellar medium (ISM: 
\citealt{spn98}), and a wind-like environment [$\rho(r) \propto r^{-2}$], 
as would be the case for a massive-star progenitor that shed its outer 
envelope at a constant rate \citep{cl00}.

At early times, the afterglow emission appears isotropic to distant observers
due to the effects of relativistic beaming.  However, the outflow slows as it
sweeps up more and more circumburst material.  When $\Gamma (t) 
\approx 1 / \theta$, lateral spreading of the jet becomes important and
observers will notice ``missing'' emission from wide angles \citep{r99,sph99}.  
This hydrodynamic transition manifests itself as an achromatic steepening
in the afterglow light curve.  Measuring the time of this jet break
($t_{\mathrm{j}}$) allows us to infer the opening angle of the outflow.

In order to ascertain the total relativistic energy output from a GRB, we 
require three measurements: (1) $E_{\gamma,\mathrm{iso}}$, the isotropic energy
release in the prompt gamma-ray emission, which is inferred from the 
high-energy fluence and the associated afterglow or host redshift; (2)
$\theta$, the half-opening angle of the beamed emission, inferred from the
detection of a jet break; and (3) $E_{\mathrm{KE}}$, the kinetic energy 
of the blast wave that is powering the broadband afterglow, which can be 
inferred either via afterglow modeling or, more accurately, from 
late-time radio calorimetry in the nonrelativistic phase 
\citep{bkf04,fsk+05,vkr+08}.  We stress here that we are neglecting 
contributions from nonelectromagnetic phenomena (neutrinos, gravity 
waves, etc.) and slower-moving material (i.e., supernova emission), and 
so are providing only lower limits on the total energy budget.

Unfortunately, none of the radio afterglows in our sample of LAT events are
sufficiently bright to perform calorimetry in the nonrelativistic phase.
Instead, we construct afterglow models (including both the standard
formulation and corrections for radiative losses and inverse-Compton
emission; \citealt{se01}) and compare these with our observations using
the multi-parameter fitting program of \citet{yhs+03}. 
Our objective is to translate the observed three critical frequencies,
together with the peak flux density, $F_{\nu,\mathrm{max}}$, and the jet break
time, $t_{\mathrm{j}}$, into a physical description of the outflow.  In
particular, we shall attempt to estimate seven parameters: $E_{\mathrm{KE}}$, the
kinetic energy of the blast wave; $n$, the particle density of the 
circumburst medium (or, alternatively for a wind-like circumburst medium, 
$A_{*}$, where $\rho = 5 \times 10^{11} A_{*} 
r^{-2}$\,g\,cm$^{-3}$)\footnote{For a wind-like medium,
the circumburst density is normalized for a progenitor mass-loss rate
of $\dot{M} = 10^{-5}$\,M$_{\odot}$\,yr$^{-1}$ and a wind speed of $v_{w} = 
1000$\,km\,s$^{-1}$ \citep{cl99}.  The units of $A_{*}$ are thus g cm$^{-1}$,
and the conversion to a particle density (as a function of radius) is 
given by $n = 30 A_{*} r_{17}^{-2}$\,cm$^{-3}$.};
$\epsilon_{e}$, the fraction of the total energy apportioned to electrons;
$\epsilon_{B}$, the fraction of the total energy apportioned to the magnetic
field; $p$, the electron power-law index; $A_{V}$, the host-galaxy 
extinction\footnote{We have assumed an SMC-like extinction curve for all
events here \citep{p92,kkz06}.  Given the relatively modest amounts of 
dust inferred for these host galaxies, this choice does not affect 
any of our primary conclusions.}; and $\theta$, the jet half-opening angle.  

Optical magnitudes have been converted to flux densities after correcting for 
Galactic extinction using zero points from 
\citet{fsi95}.  To account for differences in instrumental configurations, we 
have applied a 7\% cross-calibration uncertainty to all data points before 
calculating the models.  All reported uncertainties have been determined 
using a Monte-Carlo bootstrap analysis with 1000 trials and 
represent only statistical errors associated with the fit.  Systematic errors 
associated with model uncertainties are potentially larger and difficult to 
estimate.

%%%%%%%%%%%%%%%%%%%%%%%%%%%%%%%%%%%%%%%%%%%%%%%%%%%%%%%%%%%%%%%%%%%%%%%%%%%
\subsection{\grba}
\label{sec:0323models}
%%%%%%%%%%%%%%%%%%%%%%%%%%%%%%%%%%%%%%%%%%%%%%%%%%%%%%%%%%%%%%%%%%%%%%%%%%%

%%%%%%%%%%%%%%%%%%%%%%%%%%%%%%%%%%%%%%%%%%%%%%%%%%%%%%%%%%%%%
\begin{deluxetable*}{cccccccc}[t]
  \tabletypesize{\scriptsize}
  \tablecaption{\grba\ Afterglow Best-Fit Parameters}
  \tablecolumns{8}
  \tablewidth{0pc}
  \tablehead{\colhead{$E_{\mathrm{KE,iso}}$} & \colhead{$A_{*}$} 
             & \colhead{$\epsilon_{e}$} & \colhead{$\epsilon_{B}$} & 
             \colhead{$\theta$} & \colhead{$p$} & \colhead{$A_{V}$(host)} & 
             \colhead{$\chi^{2}_{\nu}$ (d.o.f.)} \\
             \colhead{($10^{52}$\,erg)} & \colhead{(g cm$^{-1}$)} & \colhead{($\%$)} & 
             \colhead{($\%$)} & \colhead{($^{\circ}$)} & & \colhead{(mag)} &
            }
  \startdata
  $200_{-30}^{+90}$ & $0.12_{-0.01}^{+0.02}$ & $6.8 \pm 0.5$ & 
  $0.31 \pm 0.12$ & $2.6_{-0.1}^{+0.6}$ & $2.81 \pm 0.06$ & $0.12 \pm 0.03$ 
  & 1.39 (70) \\
  \enddata
\label{tab:0323mod}
\end{deluxetable*}
%%%%%%%%%%%%%%%%%%%%%%%%%%%%%%%%%%%%%%%%%%%%%%%%%%%%%%%%%%%%

%%%%%%%%%%%%%%%%%%%%%%%%%%%%%%%%%%%%%%%%%%%%%%%%%%%%%%%%%%%%%%%%%%%%%%%%%%%
\subsubsection{Preliminary Considerations}
\label{sec:0323prelim}
Before proceeding to a detailed model, we derive some initial constraints by 
looking at the spectral and temporal behavior of the afterglow.  Considering 
first the X-ray afterglow, we fit the flux density to a power-law decay 
and find a best-fit index of $\alpha_{X} = 1.5 \pm 0.1$ ($\chi^{2} = 28.2$ for
22 degrees of freedom, d.o.f.).  Performing an analogous power-law fit to 
the X-ray spectrum, we find $\beta_{X} = 1.06_{-0.13}^{+0.34}$ ($\chi^{2} = 
8.0$ for 13 d.o.f.; see the on-line compilation of N.R.B.~for details).   

We perform a similar fit in the optical ($r^{\prime}$, $R$, and $i^{\prime}$ 
filters), forcing the decay index to be identical in all filters and ignoring 
the points at $t > 10$\,days due to host galaxy contamination (see below).  
Though the fit quality is much worse ($\chi^{2} = 178.4$ for 34 d.o.f.), we 
find a similar decay index as seen in the X-rays: $\alpha_{O} = 1.8 \pm 0.1$.  
Likewise, a spectral fit of the multi-color GROND data obtained at 
$t = 1.12$\,days results in a spectral index of $\beta_{O} = 0.95 \pm 0.01$.  

The comparable spectral and temporal indices in the optical and X-rays 
suggests that both bandpasses fall in the same synchrotron spectral regime.  In 
other words, the cooling frequency should fall either below the optical or 
above the X-ray bandpass for the duration of these observations ($t 
\approx 1$--$10$\,days).  Examining the ``closure'' relations between $\alpha$ 
and $\beta$ in different circumburst media and spectral regimes (e.g., 
\citealt{pbr+02}), we find that we can rule out a low cooling frequency at 
large significance, as this would require $\alpha = (3 \beta - 1) / 2 
\approx 1.0$ for both a wind-like and ISM-like circumburst medium.

The relatively flat radio light curve (modulo effects from interstellar 
scintillation) until $t \approx 20$\,days has two important implications.  
First, it suggests a wind-like circumburst medium, as the flux would be 
expected to rise in proportion to $t^{1/2}$ for $\nu_{a} < \nu < \nu_{m}$ in a 
constant-density environment.  From this and our previous X-ray and 
optical spectral and temporal decay indices, we can infer that the electron 
spectral index is relatively steep, $p \approx 
2.7$ (e.g., \citealt{skr06,svr+08,csv+09}).  Second, together with the lack 
of evidence for late-time steepening in the X-ray and optical light curves, 
this suggests that if a jet break is required, it should occur at $t 
\gtrsim 20$\,days. (Another possibility for the decay could be the peak 
frequency $\nu_{m}$ passing through the radio bands.)

Finally, the relatively bright $r^{\prime}$-band detection at $t = 130$\,days is 
almost certainly dominated by flux from the host galaxy and not the afterglow 
of \grba.  If we assume this flux is due entirely to host light, the host 
will contribute a significant fraction of the flux in some of the late-time 
points of our optical light curve ($\sim 40$\% at $t = 14$\,days in 
$r^{\prime}$).  We assume a flat spectrum and subtract a host contribution 
of $f_{\nu} = 0.40$\,$\mu$Jy from all optical data points for our
afterglow modeling.  
 
%%%%%%%%%%%%%%%%%%%%%%%%%%%%%%%%%%%%%%%%%%%%%%%%%%%%%%%%%%%%%%%%%%%%%%%%%%%
\subsubsection{Modeling Results}
\label{sec:0323results}
With the above constraints in hand, we have modeled the afterglow of \grba\ with
the software described above.  The resulting best-fit wind model is plotted in 
Figure~\ref{fig:0323} and the derived parameters are provided in 
Table~\ref{tab:0323mod}.  The overall fit quality is reasonable 
($\chi^{2} =  97.4$ for 70 d.o.f.), with comparable residuals in the radio, 
optical, and X-ray bandpasses.  We were unable to find any physically 
plausible models for a constant-density medium with reasonable fit quality.

As expected, the cooling frequency $\nu_{c}$ lies at or above the X-ray 
bandpass for the duration of our observations ($t \gtrsim 1$\,day).  As a 
result, the ratio of $E_{\mathrm{KE,iso}}$ to $A_{*}$ is required to be 
larger than inferred for previous GRBs (e.g., \citealt{pk01,yhs+03}).  While 
the isotropic blast-wave kinetic energy is relatively large compared to that of
previously modeled GRBs ($E_{\mathrm{KE,iso}} = 2.0 \times 10^{54}$\,erg), it 
is in fact comparable to the prompt gamma-ray energy release, as would be 
expected for reasonable values of the gamma-ray efficiency $\eta_{\gamma}$.  
The inferred density is slightly lower than usual ($A_{*} = 0.12$\,g cm$^{-1}$), 
though smaller values have been reported in the literature (e.g., GRB\,020405; 
\citealt{clf04}).  

The models result in a jet break time of $t_{\mathrm{j}} = 
17.8^{+19.6}_{-1.7}$\,days.  This occurs after the X-ray and optical 
observations have stopped (except for the host detection in the optical), 
and is therefore most directly constrained by the radio light curve.  In 
Figure~\ref{fig:0323} we also plot the same model parameters but for an 
isotropic explosion, where the turnover in the radio at late times is 
due to the peak frequency $\nu_{m}$ passing through the radio bandpass.  
The current radio data provide only weak constraints on the jet break time 
upper bound, although the models suggest a much smaller uncertainty in the 
opening angle.  We return to the robustness of our determination of $\theta$ 
in \S~\ref{sec:limit}.

While the jet break occurs rather late in the observer frame 
(e.g., \citealt{zkk06}), the inferred opening angle ($\theta = 2.6^{\circ}$) 
is still relatively small compared with previous samples.  Though 
the dependence is not strong ($\theta \propto (E_{\mathrm{KE,iso}} / 
A_{*})^{-1/4}$), the large ratio of $E_{\mathrm{KE,iso}}$ to $A_{*}$ effectively 
lowers the opening angle for a given jet break time.  

After applying the collimation correction, we find that the true energy release of 
\grba\ is $E_{\gamma} = 4.1_{-0.8}^{+2.9} \times 10^{51}$\,erg, 
$E_{\mathrm{KE}} = 2.1_{-0.5}^{+2.4} \times 10^{51}$\,erg.  We compare these 
results with a larger sample of events in \S~\ref{sec:engines}.

%%%%%%%%%%%%%%%%%%%%%%%%%%%%%%%%%%%%%%%%%%%%%%%%%%%%%%%%%%%%%%%%%%%%%%%%%%%
\subsection{\grbb}
\label{sec:0328models}
%%%%%%%%%%%%%%%%%%%%%%%%%%%%%%%%%%%%%%%%%%%%%%%%%%%%%%%%%%%%%%%%%%%%%%%%%%%

%%%%%%%%%%%%%%%%%%%%%%%%%%%%%%%%%%%%%%%%%%%%%%%%%%%%%%%%%%%%%
\begin{deluxetable*}{cccccccc}[t]
  \tabletypesize{\scriptsize}
  \tablecaption{\grbb\ Afterglow Best-Fit Parameters}
  \tablecolumns{8}
  \tablewidth{0pc}
  \tablehead{\colhead{$E_{\mathrm{KE,iso}}$} & \colhead{$A_{*}$} 
             & \colhead{$\epsilon_{e}$} & \colhead{$\epsilon_{B}$} & 
             \colhead{$\theta$} & \colhead{$p$} & \colhead{$A_{V}$(host)} & 
             \colhead{$\chi^{2}_{\nu}$ (d.o.f.)} \\
             \colhead{($10^{52}$\,erg)} & \colhead{(g cm$^{-1}$)} & \colhead{($\%$)} & 
             \colhead{($\%$)} & \colhead{($^{\circ}$)} & & \colhead{(mag)} &
            }
  \startdata
  $11_{-3}^{+5}$ & $0.11_{-0.05}^{+0.03}$ & $9 \pm 2$ & 
  $0.9_{-0.2}^{+0.7}$ & $5.2_{-0.7}^{+1.4}$ & $2.58_{-0.07}^{+0.12}$ & 
  $0.41 \pm 0.04$ & 1.63 (43) \\
  \enddata
\label{tab:0328mod}
\end{deluxetable*}
%%%%%%%%%%%%%%%%%%%%%%%%%%%%%%%%%%%%%%%%%%%%%%%%%%%%%%%%%%%%

%%%%%%%%%%%%%%%%%%%%%%%%%%%%%%%%%%%%%%%%%%%%%%%%%%%%%%%%%%%%%%%%%%%%%%%%%%%
\subsubsection{Preliminary Considerations}
\label{sec:0328prelim}
Following the results of \S~\ref{sec:0323prelim}, we first consider the
spectral and temporal indices of \grbb\ in the observed bandpasses 
independently.  The X-ray afterglow light curve is well fit by a single 
power-law decay with index $\alpha_{X} = 1.65 \pm 0.09$ ($\chi^{2} = 
24.1$ for 26 d.o.f.).  The spectral index, on the other hand, is less well 
constrained: $\beta_{X} = 1.1_{-0.3}^{+0.4}$ ($\chi^{2} = 6.7$ for 14 d.o.f.).  
We note that the X-ray spectral fit indicates the presence of an additional 
absorption component in the host galaxy of \grbb\ ($N_{H} = 4.9_{-2.8}^{+3.5} 
\times 10^{21}$\,cm$^{-2}$ at $z = 0.736$).

The relative paucity of optical data, particularly at late times, makes 
a similar analysis challenging.  Fitting the UVOT $U$-band light curve to a
single power-law decay results in an index of $\alpha_{O} = 1.2 \pm 0.1$;
however, the quality of the fit is extremely poor ($\chi^{2} = 24.5$ for 5 
d.o.f.).  An initially steep decay in the light curve becomes more shallow 
at later times, much like is seen in the UVOT white filter \citep{GCNR.207}.  
This result suggests the presence of host-galaxy contamination at late times.

The best-fit spectral index assuming a power-law model for the simultaneous 
multi-color GROND data is $\beta_{O} = 1.24 \pm 0.04$, 
although the fit quality is extremely poor ($\chi^{2} = 85.5$ for 5 d.o.f.).  
The poor fit statistic is dominated almost entirely by a single filter 
($J$ band), however, and the relatively steep spectral index (compared to 
$\beta_{OX} \approx 0.8$) appears to be robust.  This is consistent with the 
presence of dust indicated by the X-ray spectral fits.

In the case of \grbb, examining the $\alpha$--$\beta$ closure relations yields
little insight.  Because of the poor fit quality in the optical and the large
uncertainty in the X-ray spectral index, we are unable to rule out much of
the available parameter space.  We note, however, that with the exception of 
the excess absorption, the X-ray afterglow light curve and spectrum of \grbb\
closely resemble those of \grba.  Similarly, if we assume that the late-time optical
data are dominated by host-galaxy light, the optical light curve is also 
compatible with that seen from \grba.

%%%%%%%%%%%%%%%%%%%%%%%%%%%%%%%%%%%%%%%%%%%%%%%%%%%%%%%%%%%%%%%%%%%%%%%%%%
\subsubsection{Modeling Results}
\label{sec:0328results}
The results of our best-fit model for the afterglow of \grbb\ are shown
in Figure~\ref{fig:0328} and the parameters are provided in 
Table~\ref{tab:0328mod}.  The overall fit quality is reasonable 
($\chi^{2} = 70.1$ for 43 d.o.f.).  Only the early radio light curve, 
which suffers from strong scintillation, differs systematically from the 
model predictions.

Much like in the case of \grba, we find that only a wind-like circumburst medium 
can reproduce the observed light curves with plausible physical parameters.
While the slow rise in the early radio afterglow is better explained by 
a constant-density environment, our modeling indicates ISM-like 
solutions require $\epsilon_{e} \rightarrow 1$ and $\epsilon_{B} \rightarrow 0$
(i.e., a fully radiative solution).

We note that our solutions require the presence of a relatively bright host
galaxy --- if the late-time $U$-band points are not predominantly due to 
host light, the fit quality of our models decreases somewhat ($\chi^{2}_{\nu}
\gtrsim 2.0$).  

The steep decline in the radio light curve at late times offers relatively
tight constraints on the presence of a jet break.  In fact, the break
is consistent with (though not formally required by) the final few points
in the X-ray light curve as well.  For our best-fit model, we infer
$t_{\mathrm{j}} = 6.4_{-1.5}^{+12.0}$\,days and $\theta = 
(5.2_{-0.7}^{+1.4})^{\circ}$.  Primarily because of the relatively sparse 
optical data, we find several unique models with similar though slightly 
worse overall fit statistics.  We note, however, that all these solutions 
result in comparable opening angles, the most important parameter for our 
energetics calculations.

Correcting for the effects of beaming, we find an energy release of 
$E_{\gamma} = 5.5_{-1.8}^{+4.3} \times 10^{50}$\,erg and $E_{\mathrm{KE}} = 
4.5_{-2.0}^{+6.1} \times 10^{50}$\,erg.  The total relativistic energy release, 
$E_{\mathrm{tot}} \approx 10^{51}$\,erg, is somewhat less than the 
value we derive for \grba.

%%%%%%%%%%%%%%%%%%%%%%%%%%%%%%%%%%%%%%%%%%%%%%%%%%%%%%%%%%%%%%%%%%%%%%%%%%
\subsection{\grbc}
\label{sec:0902models}
%%%%%%%%%%%%%%%%%%%%%%%%%%%%%%%%%%%%%%%%%%%%%%%%%%%%%%%%%%%%%%%%%%%%%%%%%%

%%%%%%%%%%%%%%%%%%%%%%%%%%%%%%%%%%%%%%%%%%%%%%%%%%%%%%%%%%%%%
\begin{deluxetable*}{cccccccc}[t]
  \tabletypesize{\scriptsize}
  \tablecaption{\grbc\ Afterglow Best-Fit Parameters}
  \tablecolumns{8}
  \tablewidth{0pc}
  \tablehead{\colhead{$E_{\mathrm{KE,iso}}$} & \colhead{$n$} 
             & \colhead{$\epsilon_{e}$} & \colhead{$\epsilon_{B}$} & 
             \colhead{$\theta$} & \colhead{$p$} & \colhead{$A_{V}$(host)} & 
             \colhead{$\chi^{2}_{\nu}$ (d.o.f.)} \\
             \colhead{($10^{52}$\,erg)} & \colhead{(cm$^{-3}$)} &
             \colhead{($\%$)} & \colhead{($\%$)} & \colhead{($^{\circ}$)} & 
             & \colhead{(mag)} &
            }
  \startdata
  $68_{-6}^{+14}$ & $(5.8_{-1.8}^{+1.2}) \times 10^{-4}$ & $15_{-1}^{+4}$ & $5.8_{-1
.0}^{+2.5}$ &
  $3.4_{-0.3}^{+0.4}$ & $2.22_{-0.04}^{+0.08}$ & $0.18_{-0.05}^{+0.08}$ &
  1.06 (137) \\
  \enddata
\label{tab:0902mod}
\end{deluxetable*}
%%%%%%%%%%%%%%%%%%%%%%%%%%%%%%%%%%%%%%%%%%%%%%%%%%%%%%%%%%%%

%%%%%%%%%%%%%%%%%%%%%%%%%%%%%%%%%%%%%%%%%%%%%%%%%%%%%%%%%%%%%%%%%%%%%%%%%%
\subsubsection{Preliminary Considerations}
\label{sec:0902prelim}
A single power-law model with $\alpha_{X} = 1.36 \pm 0.03$ provides a good
fit to the X-ray light curve of \grbc\ over the entire span of the XRT
observations ($\chi^{2} = 94.8$ for 107 d.o.f.).  Likewise, the X-ray 
spectrum is well fit by a power law with $\beta_{X} = 0.90 \pm 0.13$
($\chi^{2} = 42.8$ for 56 d.o.f.).  The X-ray spectrum strongly favors
the presence of absorption in excess of the Galactic value along the 
GRB line of sight.  Assuming the dust arises in the GRB host galaxy
at $z = 1.82$, the best-fit host column density is $N_{H} = 
(6.3_{-1.7}^{+2.1}) \times 10^{21}$\,cm$^{-2}$.

The decline inferred from the early-time ROTSE-III $R$-band observations 
is significantly steeper than the late-time ($t \gtrsim 1$\,day) optical
decay.  Following \citet{psp+10}, we attribute this emission largely
to the presence of a reverse shock and do not include
these data points in any subsequent analysis.  Simultaneously 
fitting the remaining optical data to a power law of the form $f_{\nu} 
\propto t^{-\alpha} \nu^{-\beta}$, we find $\alpha_{O} = 0.89 \pm 0.05$
and $\beta_{O} = 0.76 \pm 0.07$ ($\chi^{2} = 36.4$ for 18 d.o.f.).

The large discrepancy between the X-ray and optical temporal decay 
slopes suggests that the two bandpasses fall in different synchrotron
spectral regimes (i.e., $\nu_{O} < \nu_{c} < \nu_{X}$).  Furthermore,
the relatively shallow optical decline strongly rules out a wind-like
circumburst medium.  We find that both the X-ray and optical bandpasses
are broadly consistent with expansion into a constant-density medium
and an electron index $p \approx 2.2$.  The observed X-ray temporal 
decline is somewhat steeper than what is predicted, but this 
could be accounted for by synchrotron radiative losses at later
times.

The radio light curve appears to decay from the earliest observations
at $t_{0} + 1.5$\,days onward.  This behavior is not expected in the simple
forward-shock model where radio afterglows are typically observed to
rise (for a constant-density medium) or remain flat (for a stratified
environment) until either a jet break occurs or the peak frequency
reaches the radio bandpass (typically weeks after the burst).

However, bright, early-time radio emission is actually rather common,
having been detected in a quarter of all radio afterglows to date
\citep{sr03,c10}.  This prompt component is most commonly attributed
to the reverse shock propagating in the shocked ejecta. The reverse-shock 
emission fades away quite quickly, and at late times ($t \gtrsim
5$\,days) the afterglow will be dominated by the forward-shock emission.
GRB\,990123 was the first known example \citep{kfs+99}, but there have
been many more claimed detections since (e.g., \citealt{bsf+03,p05a,cff+10}).

The radio light curve of GRB\,090902B most closely resembles that of 
GRB\,991216 in which the early power-law decline is due to the overlap 
of emission from the reverse and forward shocks \citep{fbg+00}. 
Since we are concerned here with the modeling of the behavior of the forward 
shock, we will exclude the early radio point and postpone a full discussion 
to a subsequent paper dealing with a compilation all known GRBs with 
prompt radio emission \citep{c10}.

%%%%%%%%%%%%%%%%%%%%%%%%%%%%%%%%%%%%%%%%%%%%%%%%%%%%%%%%%%%%%%%%%%%%%%%%%%
\subsubsection{Modeling Results}
\label{sec:0902results}
Using the constraints derived above, our best-fit model for \grbc\
assuming a constant-density circumburst medium is shown in 
Figure~\ref{fig:0902}, and the derived model parameters are presented in 
Table~\ref{tab:0902mod}.  The overall fit quality is quite reasonable 
($\chi^{2} = 145.8$ for 137 d.o.f.), although the fit quality is 
somewhat better in the X-ray and radio than in the optical/UV.  We stress, 
however, that we have not included the first two $R$-band data points and 
the first VLA detection in our fitting, as we believe the flux at these 
points is dominated by reverse-shock emission.

Much like the other events considered here, our inferred parameters for
\grbc\ suggest a large afterglow kinetic energy ($E_{\mathrm{KE,iso}} = 
6.8_{-0.6}^{+1.4} \times 10^{53}$\,erg; a factor of five less than the 
prompt gamma-ray energy release) and a low circumburst density 
($n = 5.8_{-1.8}^{+1.2} \times
10^{-4}$\,cm$^{-3}$).  In fact, we are unable to find any acceptable
solutions ($\chi^{2}_{\nu} < 2$) with $n > 10^{-2}$\,cm$^{-3}$.
The inferred circumburst density for many 
long-duration GRBs is lower than typical values observed in dense
molecular clouds (where presumably their massive-star progenitors have
formed).  Yet the value we have derived for \grbc\ is closer in fact
to what one might find in the ambient ISM or even the intergalactic 
medium (IGM), as has been found for the few short-hard GRBs with
sufficient afterglow data (e.g., \citealt{p06,pmg+09}).  We return to
this issue in more detail in \S~\ref{sec:comp}.

Neither the X-ray nor the (less constraining) optical bandpass
formally require a jet break over the duration of our observations.  In this
case, like \grba, we must rely largely on the radio.  Our models suggest
that the peak synchrotron frequency, $\nu_{m}$, will not fall to the radio
bandpass until $t \gtrsim 40$\,days.  The declining radio light curve
at late times therefore requires a jet break at $t_{\mathrm{j}} = 
6.2_{-0.8}^{+2.4}$\,days.  Again, because of the large ratio of 
$E_{\mathrm{KE,iso}}$ to $n$, this corresponds to a relatively narrow
opening angle: $\theta = (3.4_{-0.3}^{+0.4})^{\circ}$.  

Correcting for collimation, we find the true energy release from
\grbc\ to be $E_{\gamma} = (5.6 \pm 1.5) \times 10^{51}$\,erg,
$E_{\mathrm{KE}} = 1.2_{-0.4}^{+0.7} \times 10^{51}$\,erg.  

%%%%%%%%%%%%%%%%%%%%%%%%%%%%%%%%%%%%%%%%%%%%%%%%%%%%%%%%%%%%%%%%%%%%%%%%%%
\subsection{\grbd}
\label{sec:0926models}
%%%%%%%%%%%%%%%%%%%%%%%%%%%%%%%%%%%%%%%%%%%%%%%%%%%%%%%%%%%%%%%%%%%%%%%%%%

%%%%%%%%%%%%%%%%%%%%%%%%%%%%%%%%%%%%%%%%%%%%%%%%%%%%%%%%%%%%%
\begin{deluxetable*}{cccccccc}[t]
  \tabletypesize{\scriptsize}
  \tablecaption{\grbd\ Afterglow Best-Fit Parameters}
  \tablecolumns{8}
  \tablewidth{0pc}
  \tablehead{\colhead{$E_{\mathrm{KE,iso}}$} & \colhead{$A_{*}$} 
             & \colhead{$\epsilon_{e}$} & \colhead{$\epsilon_{B}$} & 
             \colhead{$\theta$} & \colhead{$p$} & \colhead{$A_{V}$(host)} & 
             \colhead{$\chi^{2}_{\nu}$ (d.o.f.)} \\
             \colhead{($10^{52}$\,erg)} & \colhead{(g\,cm$^{-1}$)} &
             \colhead{($\%$)} & \colhead{($\%$)} & \colhead{($^{\circ}$)} & 
             & \colhead{(mag)} &
            }
  \startdata
  $24.2_{-1.3}^{+1.9}$ & $0.67_{-0.02}^{+0.14}$ & $30_{-3}^{+2}$ & 
  $0.87_{-0.03}^{+0.07}$ & $7.1 \pm 0.2$ & $2.19 \pm 0.02$ & $0.06 \pm 0.01$ &
  1.18 (149) \\
  \enddata
\label{tab:0926mod}
\end{deluxetable*}
%%%%%%%%%%%%%%%%%%%%%%%%%%%%%%%%%%%%%%%%%%%%%%%%%%%%%%%%%%%%

%%%%%%%%%%%%%%%%%%%%%%%%%%%%%%%%%%%%%%%%%%%%%%%%%%%%%%%%%%%%%%%%%%%%%%%%%%
\subsubsection{Preliminary Considerations}
\label{sec:0926prelim}
Fitting a single power-law decay to the X-ray light curve, we find a reasonable
quality fit ($\chi^{2} = 120.8$ for 97 d.o.f.) with $\alpha_{X} = 1.43 \pm 
0.03$.  The fit quality improves somewhat if we allow for a steeper decay at 
late times, although the post-break decay index is not particularly well 
constrained: $\alpha_{X,1} = 1.35 \pm 0.04$, $\alpha_{X,2} = 
2.6_{-0.5}^{+0.9}$, $t_{b} = 9.1_{-1.4}^{+2.1}$\,days, $\chi^{2} = 111.9$ 
(95 d.o.f.).  The X-ray spectrum is well fit ($\chi^{2} = 41.7$ for 51 
d.o.f.) by a single power law with index $\beta_{X} = 1.12 \pm 0.13$ and 
does not require any absorption in addition to the Galactic component.  

The behavior in the optical bandpass, however, is significantly more
complex (Fig.~\ref{fig:0926opt}).  The flux in the initial UVOT $U$-
and $V$-band observations declines until $t \approx 0.8$\,day, at which
point a rebrightening or plateau phase is evident in all five observed
filters ($UBVRI$).  The optical light curve of \grbd\ is reminiscent
of the rebrightening seen at $t \approx 1$\,day from GRB\,970508 
\citep{dmk+97} and GRB\,071003 (\citealt{plc+08} and references
therein).  Such a
phenomenon is difficult to reconcile with the standard afterglow
paradigm.  Possible explanations include a sharp change in the circumburst 
density (e.g., \citealt{lrc+02,tph+05} but, c.f.~\citealt{ng07}) or a smooth
injection of energy into the forward shock from the central engine
(e.g., \citealt{rm98}).  A more detailed analysis of the early-time
behavior of \grbd\ is beyond the scope of this work.

The optical light curve appears to peak at $t \approx 1$\,day in all bands.  
Fitting the PROMPT $BVRI$ data at $t > 1$\,day to a single power law results 
in a best-fit index of $\alpha_{O} = 1.38 \pm 0.02$ ($\chi^{2} = 181.6$ for 
119 d.o.f.).  However, extrapolating these results to the late-time Gemini 
imaging at $t \approx 23$\,days greatly overestimates the observed flux 
(Fig.~\ref{fig:0926late}).  Much like the X-ray data, this strongly suggests a 
steepening of the optical light curve at $t \gtrsim 8$\,d.  

We have also performed a joint spectral and temporal fit to the PROMPT data
to estimate the optical spectral index, $\beta_{O}$.  We include only the 
$V$-, $R$-, and $I$-band data, as both the $U$ and $B$ bands are likely affected
by Ly$\alpha$ absorption at $z = 2.11$.  We then find a best-fit optical
spectral index of $\beta_{O} = 1.03 \pm 0.05$.  The formal fit quality is 
relatively poor, however ($\chi^{2} = 183.2$ for 68 d.o.f.).  Finally, the
optical to X-ray spectral index at $t \gtrsim 1$\,day is $\beta_{OX} 
\approx 1.1$.

Taken together, we find similar temporal ($\alpha_{X,1} = 1.35$, $\alpha_{O,1} = 
1.38$) and spectral ($\beta_{X} = 1.12$, $\beta_{O} = 1.03$) indices in the 
X-ray and optical for $t \approx$ 1--9\,days.  Furthermore, the optical to 
X-ray spectral index, $\beta_{OX}$, is comparable to both the optical and 
X-ray spectral indices.  These facts strongly suggest that both the X-ray 
and optical bandpasses fall in the same synchrotron spectral regime.

Considering the various synchrotron closure relations, the best fit appears to 
occur when both bandpasses fall above the cooling frequency, $\nu_{c}$.  The
afterglow decay in this regime is independent of circumburst medium.  Such 
a low cooling frequency is somewhat unusual, though not unprecedented
(e.g., GRB\,050904; \citealt{fck+06}) in GRB afterglows.  The implied 
electron index in this case would be $p \approx 2.3$.

Finally, we note that the steepening in both the X-ray and optical bands at
$t \approx 9$\,days is strongly suggestive of a jet break.  We examine this 
possibility in greater detail in the following section.  

%%%%%%%%%%%%%%%%%%%%%%%%%%%%%%%%%%%%%%%%%%%%%%%%%%%%%%%%%%%%%%%%%%%%%%%%%%
\subsubsection{Modeling Results}
\label{sec:0926results}
Unfortunately, without a radio light curve we cannot uniquely solve for the
physical parameters of the afterglow of \grbd.  We can, however,
at the very least more robustly constrain the jet break time and opening angle
of this event, as well as identify broad trends in the parameters associated 
with acceptable solutions.  As discussed previously, we consider only data
at $t \ge 1$\,day due to the rebrightening observed in the optical before 
then.

As expected, we were able to find a number of acceptable solutions, both for 
constant-density and wind-like environments.  One example for a wind-like
circumburst environment is shown in Table~\ref{tab:0926mod} and 
Figure~\ref{fig:0926late}.  The model provides a reasonable description of the
data in all bandpasses, with $\chi^{2} = 175.1$ (149 d.o.f.).  Without
radio data, however, the solution presented in Table~\ref{tab:0926mod}
is not unique in a global sense, and therefore should only be taken as
representative of a larger family of solutions.

Most importantly for our purposes, a jet break at $t \approx 9 \pm 2$\,days is 
required to provide a reasonable fit to the late-time X-ray and (particularly)
optical data (Fig.~\ref{fig:0926late}).  The precise translation into an 
opening angle is dependent on the circumburst density (both profile and 
absolute value) and the afterglow energy.  Taking the values derived above
into account, we find a typical value for a wind-like circumburst
medium of $\theta \approx 7^{\circ}$.  Based on 
the observed spread of $E_{\mathrm{KE}}$ and $A_{*}$ in our wind models, 
we adopt an approximate uncertainty in this value of $\theta 
\approx (7_{-1}^{+3})^{\circ}$.

Correcting for collimation, we find a prompt energy release of $E_{\gamma} = 
1.4_{-0.4}^{+1.5} \times 10^{52}$\,erg.  Assuming a reasonable value
for the gamma-ray efficiency ($\eta_{\gamma} \gtrsim 10$\%), we adopt
a lower limit on the isotropic blast-wave energy of $E_{\mathrm{KE,iso}} 
\gtrsim 10^{53}$\,erg.  We then find a collimation-corrected 
afterglow energy of $E_{\mathrm{KE}} \gtrsim 5 \times 10^{50}$\,erg.

%%%%%%%%%%%%%%%%%%%%%%%%%%%%%%%%%%%%%%%%%%%%%%%%%%%%%%%%%%%%%%%%%%%%%%%%
\section{Discussion}
\label{sec:discussion}
%%%%%%%%%%%%%%%%%%%%%%%%%%%%%%%%%%%%%%%%%%%%%%%%%%%%%%%%%%%%%%%%%%%%%%%%

%%%%%%%%%%%%%%%%%%%%%%%%%%%%%%%%%%%%%%%%%%%%%%%%%%%%%%%%%%%%%
\begin{deluxetable*}{lcccccccc}[t]
  \tabletypesize{\scriptsize}
  \tablecaption{Summary of GRB Parameters}
  \tablecolumns{9}
  \tablewidth{0pc}
  \tablehead{\colhead{GRB} & \colhead{$z$} & \colhead{$f_{\gamma}$\tablenotemark{a}} & 
                    \colhead{$E_{\gamma,\mathrm{iso}}$\tablenotemark{a}} & \colhead{$\Gamma_{0}$} &
                    \colhead{$\theta$} & \colhead{$E_{\gamma}$} & 
                    \colhead{$E_{\mathrm{KE}}$} & \colhead{$n / A_{*}$} \\
                    & & \colhead{($10^{-4}$\,erg\,cm$^{-2}$)} & \colhead{($10^{54}$\,erg)} & &
                    \colhead{($^{\circ}$)} &
                    \colhead{($10^{51}$\,erg)} & 
                    \colhead{($10^{51}$\,erg)} &
                    \colhead{(cm$^{-3}$ / g\,cm$^{-1}$)} }
  \startdata
  090323 & 3.568 & $1.50 \pm 0.20$ & $3.99 \pm 0.53$ & $\gtrsim 600$ & $2.6_{-0.1}^{
+0.6}$ &
    $4.1_{-0.8}^{+2.9}$ & $2.1_{-0.5}^{+2.4}$ & $0.12_{-0.01}^{+0.02}$ \\
  090328 & 0.7357 & $0.73 \pm 0.08$ & $0.134 \pm 0.014$ & $ \gtrsim 200$ &
    $5.2_{-0.7}^{+1.4}$ & $0.55_{-0.18}^{+0.43}$ & $0.45_{-0.20}^{+0.61}$ & $0.11_{-
0.05}^{+0.03}$ \\
  090902B & 1.8229 & $3.83 \pm 0.05$ & $3.20 \pm 0.04$ & $\gtrsim 1000$ &
    $3.4_{-0.3}^{+0.4}$ & $5.6 \pm 1.5$ & $1.2_{-0.4}^{+0.7}$ & $(5.8_{-1.8}^{+1.2})
 \times 10^{-4}$ \\
  090926A & 2.1062 & $1.73 \pm 0.03$ & $1.89 \pm 0.03$ & $\gtrsim 700$ &
    $7^{+3}_{-1}$ & $14_{-4}^{+15}$ & $\gtrsim 0.5$ & $\cdots$\tablenotemark{b} \\
 \enddata
  \tablenotetext{a}{1--10$^{4}$\,keV observer frame bandpass.}
  \tablenotetext{b}{We have not included an estimate for the density of \grbd, as this
    parameter was not well constrained due to the lack of radio data.}
\label{tab:summary}
\end{deluxetable*}
%%%%%%%%%%%%%%%%%%%%%%%%%%%%%%%%%%%%%%%%%%%%%%%%%%%%%%%%%%%%%

%%%%%%%%%%%%%%%%%%%%%%%%%%%%%%%%%%%%%%%%%%%%%%%%%%%%%%%%%%%%%%%%%%%%%%
\subsection{Central-Engine Constraints I: Energetics and Remnants}
\label{sec:engines}
%%%%%%%%%%%%%%%%%%%%%%%%%%%%%%%%%%%%%%%%%%%%%%%%%%%%%%%%%%%%%%%%%%%%%%
In Table~\ref{tab:summary}, we summarize the primary results from this 
work, including the redshift, initial Lorentz factor, beaming angle, 
density, and collimation-corrected energy release for each of the four LAT 
GRBs considered here.

Before the launch of the \swift\ satellite in 2004, the majority of all 
well-observed afterglows were inferred to be highly collimated, with
opening angles $\theta \lesssim 10^{\circ}$ \citep{zkk06}.  The most
notable exceptions were the handful of the most nearby events, including
GRB\,980425 \citep{gvv+98,kfw+98}, GRB\,031203 
\citep{skb+04,sls04,thw+04,cbv+04,mtc+04,gmf+04}, 
and (post-\swift) GRB\,060218 
\citep{mha+06,cmb+06,skn+06,msg+06,sjf+06,pmm+06,fkz+06,mkt+07}, 
all of which appear to be isotropic explosions that were energetically 
dominated by their nonrelativistic ejecta (i.e., their associated 
supernovae).  

The typical GRB afterglows discovered by \swift, however, did not fit neatly
into this simple bimodal picture.  The afterglows of most \swift\
GRBs, including the large fraction at cosmological distances, exhibit
a much broader range of opening angles \citep{kb08,rlb+09}, with some
extreme events lacking a detectable jet break signature in the X-rays
out to hundreds of days after the burst (e.g., GRB\,060729; 
\citealt{ggw+07,gbw+10}).

All four of the LAT-detected events we have studied here are consistent 
with a relatively high degree of collimation ($\theta \lesssim 10^{\circ}$).
In this respect, then, the afterglows of LAT-detected GRBs 
more closely resemble the pre-\swift\ sample.  This in and of itself is 
not entirely surprising, given the tremendous high-$E_{\gamma,\mathrm{iso}}$
bias for events detected by the LAT (Figure~\ref{fig:egamma}).

\begin{figure*}[t]
\epsscale{1.1}
\plotone{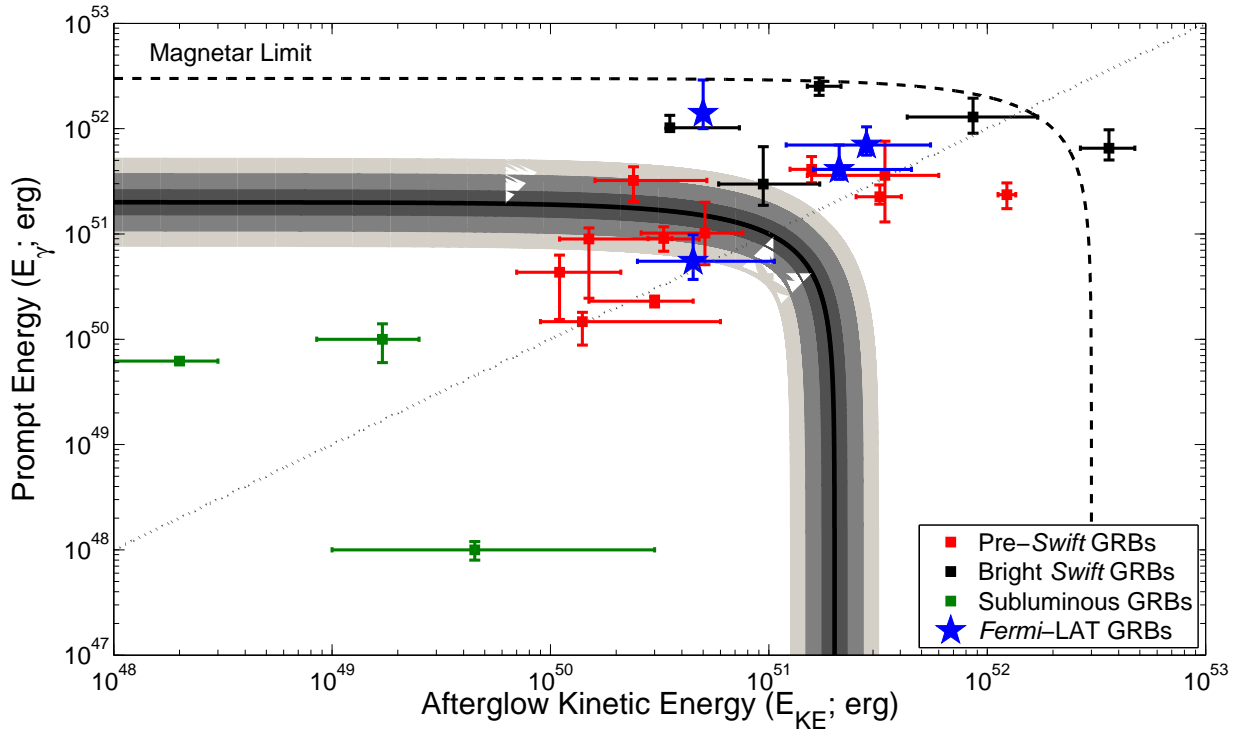}
%\plotone{../Figures/waterfall_lat.eps}
\caption{Two-dimensional relativistic energy release ($E_{\mathrm{rel}} \approx
E_{\gamma} + E_{\mathrm{KE}}$) from GRBs.  Cosmologically distant ($z \gtrsim 0.5$)
events from the pre-\swift\ era are shown in red.  The logarithmic mean
for these events, $\langle E_{\mathrm{rel}} \rangle = 2 \times 10^{51}$\,erg, is 
indicated by the solid black line.  Shaded regions correspond to 1$\sigma$, 2$\sigma$, 
and 3$\sigma$ errors on this mean value.  The three most nearby events (GRBs\,980425, 031203,
and 060218) are plotted in green and are underluminous by several orders of 
magnitude.  The four LAT events from this work are plotted in blue; all but 
\grbb\ fall at the high end of the pre-\swift\ distribution (note that we have not
plotted horizontal errors bars for \grbd\ due to the large uncertainty in $E_{\mathrm{KE}}$).
Instead, they are more consistent with some of the brightest events from the
\swift\ era (black squares).  The total relativistic energy release from \grbd\ 
appears to exceed $10^{52}$\,erg.  Such
hyper-energetic events pose a severe challenge to the magnetar models,
where the total energy release cannot exceed $3 \times 10^{52}$\,erg (dashed
black line).  References --- \citet{pk02}: 
GRBs\,990123, 990510, 991208, 991216, 000301C, 010222;
\citet{yhs+03}: GRBs\,970508, 980703, 000926; \citet{bkf04}: GRBs\,970508,
980703; \citet{clf04}: GRB\,020405; \citet{bdf+01}: GRB\,000418; \citet{lc99}:
GRB\,980425; \citet{skb+04}: GRB\,031203; \citet{skn+06}: GRB\,060218; 
\citet{fck+06}: GRB\,050904; \citet{ccf+08}: GRB\,070125; \citet{cfh+10}:
GRBs\,050820A, 060418, 080319B. }
\label{fig:energy}
\end{figure*}

In Figure~\ref{fig:energy}, we plot the two-dimensional (prompt +
afterglow) collimation-corrected relativistic energy release for the 
four LAT events in this work, compared with several additional samples.
Shown in red are 11 pre-\swift\ GRBs at cosmological distances ($z > 0.5$)
for which sufficient broadband (X-ray, optical, and radio) data exist to
constrain both $E_{\gamma}$ and $E_{\mathrm{KE}}$.  We find the logarithmic
mean for the sum of these two values, $E_{\mathrm{rel}} \approx E_{\gamma} 
+ E_{\mathrm{KE}}$,\footnote{Assuming an approximately log-normal distribution,
we calculate the weighted mean of $\langle \log_{10}(E_{\mathrm{tot}}) \rangle$, 
where $\sigma = \delta E_{\mathrm{tot}} / (\ln(10) \times  E_{\mathrm{tot}})$.}
to be $\langle E_{\mathrm{rel}} \rangle = 2.0 \times
10^{51}$\,erg, with an error in the mean of 0.17 dex.  The solid black
line indicates a constant value of $E_{\mathrm{rel}}$ corresponding to this
mean, while the gray shaded regions indicate 1$\sigma$, 2$\sigma$, and 3$\sigma$
confidence intervals.

The gray dotted line in Figure~\ref{fig:energy} represents a constant
gamma-ray efficiency of $\eta_{\gamma} = 50$\%.  Most events are
roughly consistent (or even exceed) this value.  This result presents
a problem for most internal shock models of the prompt emission, which
predict a maximal gamma-ray efficiency of $\eta_{\gamma} \lesssim
10$\% \citep{kps97,dm98}.  Alternatively, if the early afterglow
undergoes a relatively long-lived radiative phase, we may be
significantly underestimating the blastwave kinetic energy \citep{ggn+10}.

As discussed previously, the three most nearby GRBs are clearly 
subenergetic outliers.  The energy release from the supernovae associated
with these events, $E_{\mathrm{SN}} \sim 10^{51}$--$10^{52}$\,erg 
\citep{imn+98,mdn+06},
is several orders of magnitude larger than the energy apportioned to
the relativistic ejecta.  To further distinguish these events, the rate of 
such subluminous GRBs suggests they are many times more common (per unit 
volume) than the typical cosmological GRBs \citep{skn+06,cbv+06,gd07}, 
although \citet{bbp09} have suggested that both samples may be described by 
a single luminosity function --- that is, subluminous GRBs may simply extend
continuously from the higher-energy population.

Of the LAT events studied here, the total relativistic energy output
from only a single one (\grbb) is consistent with the pre-\swift\ mean
at the 3$\sigma$ level.  This is not entirely surprising, however, as
the $E_{\mathrm{rel}}$ distribution of pre-\swift\ GRBs exhibits a 
reasonable dispersion ($\sim 0.55$\,dex, or a factor of 3.5).  It is
clear, however, that \grba, \grbc, and \grbd\ all fall at the very 
high end of the pre-\swift\ distribution.  In particular, the only 
event comparable to \grbd, with $E_{\gamma} \approx 10^{52}$\,erg and 
$E_{\mathrm{KE}}$ relatively unconstrained, is GRB\,970508 ($E_{\mathrm{rel}}
\approx 1.5 \times 10^{52}$\,erg, dominated by the large afterglow
kinetic energy; \citealt{yhs+03,bkf04}).

GRBs\,090323, 090902B, and 090926A appear consistent instead with 
a subsample of the brightest
\swift\ events from \citet{cfh+10} (see also \citealt{fck+06,ccf+08}).
This, too, is not unexpected, as these \swift\ events were chosen on the
basis of large $E_{\gamma,\mathrm{iso}}$ values, and therefore should
in large part mimic at least this component of the LAT selection
effects.

Much like several events in the bright \swift\ sample, \grbd\ appears 
to exceed the canonical GRB relativistic energy release of $10^{51}$\,erg
by roughly an order of magnitude.  We stress that our methodology 
provides a relatively conservative estimate for the prompt energy
release of \grbd, for a number of reasons.  First, we consider only
the rest-frame 1--10$^{4}$\,keV bandpass for \textit{all} events considered
in Figure~\ref{fig:energy}.  Extrapolating measurements from previous
instruments, with bandpasses typically extending only to $\sim 1$\,MeV
(only a few hundred keV for \swift), up to the GeV range would introduce
significant uncertainties in $E_{\gamma,\mathrm{iso}}$.  While allowing for a
more robust burst-to-burst comparison, we have not included a significant
fraction of the detected high-energy fluence in our energy calculations
for these LAT events ($\sim 30\%$ for \grbd).  

Second, we have derived the opening angles of all our events using 
parameters inferred from our broadband modeling.  Consequently, the 
$E_{\mathrm{rel}}$ we derive for \grbc\ is a factor of several lower
than that reported by other authors (\S~\ref{sec:comp} and \ref{sec:limit}).
Most importantly, for \grbd\ we have deliberately used only wind-like
models to derive the beaming angle, even though our data are insufficient
to rule out a constant-density medium.  For ISM-like models, the
opening angles we found were on average $\sim 75\%$ larger, corresponding
to a beaming-corrected prompt energy release of $E_{\gamma} \approx 4 \times
10^{52}$\,erg.

Together with the discovery of several comparable events to \grbd\ 
in the past few years, including GRB\,050904 ($E_{\mathrm{rel}} \approx
2 \times 10^{52}$\,erg; \citealt{fck+06}), GRB\,070125 ($E_{\mathrm{rel}} 
\approx 3 \times 10^{52}$\,erg; \citealt{ccf+08}), GRB\,050820A
($E_{\mathrm{rel}} \approx 4 \times 10^{52}$\,erg; \citealt{cfh+10}), and
GRB\,090423 ($E_{\mathrm{rel}} \gtrsim 1 \times 10^{52}$\,erg;
\citealt{cff+10}), we now believe there is substantial evidence in favor of
a subpopulation of GRBs with relativistic energy outputs either very near 
or above $10^{52}$\,erg.  We refer to such events as hyper-energetic GRBs 
in what follows, and outline some of the implications of this particular
energy threshold.

Much as was argued by \citet{srv+09} for the case of GRB\,080721,
the total energy budget is an important diagnostic for any central-engine 
model.  In particular, models for which the outflow is powered
by the spin-down of a highly magnetized ($B \gtrsim 10^{15}$\,G)
proto-neutron star are subject to strict constraints on the total 
energy budget of $E_{\mathrm{total}} < 3 \times 10^{52}$\,erg (the 
rotational energy of a maximally spinning 1.4\,M$_{\odot}$ neutron
star; e.g., \citealt{tcq04,mtq07}).  

Only accounting for the relativistic 
energy output, \grbd\ seems to approach within at least
a factor of a few of this limit.  While we caution that there are still 
significant uncertainties associated with the models used to infer 
the afterglow parameters (and hence $E_{\mathrm{rel}}$), we have not
yet accounted for additional sources of energy, including (nonrelativistic)
SN emission, radiative losses at early times due to bright X-ray
flares \citep{brf+05,fmr+07}, and synchrotron losses during the later phases
of afterglow evolution \citep{yhs+03}.  Even if we have overestimated
the relativistic energy output for these events by a factor of 
several and the magnetar energy limit is not strictly violated, 
the tremendous efficiency required by this process strains
credulity. 

%%%%%%%%%%%%%%%%%%%%%%%%%%%%%%%%%%%%%%%%%%%%%%%%%%%%%%%%%%%
%%%%%%%%%%%%%%%%%%%%%%%%%%%%%%%%%%%%%%%%%%%%%%%%%%%%%%%%%%%%

%%%%%%%%%%%%%%%%%%%%%%%%%%%%%%%%%%%%%%%%%%%%%%%%%%%%%%%%%%%%%%%%%%%%%%
\subsection{Central-Engine Constraints II: Lorentz Factor, Opening Angle, and
                     Acceleration Mechanisms}
\label{sec:acceleration}
%%%%%%%%%%%%%%%%%%%%%%%%%%%%%%%%%%%%%%%%%%%%%%%%%%%%%%%%%%%%%%%%%%%%%%
Unlike magnetar models that are powered by the spin-down of a highly 
magnetized proto-neutron star, GRB models in which the core
of an evolved massive star collapses promptly to form a black
hole and accretion disk system (``Type I collapsars'' following the 
nomenclature of \citealt{mw01}) have significantly relaxed constraints on
the total energy budget that are in at least some cases capable of 
accommodating hyper-energetic events.  For instance, the initial models
of \citet{mw99} that begin with a 35\,M$_{\odot}$ He star with a 
10\,M$_{\odot}$ evolved core lead to accretion rates of $\sim 
0.1$\,M$_{\odot}$\,s$^{-1}$ onto a 3\,M$_{\odot}$ rotating black hole.
Assuming approximately continuous feeding over the fallback time
of the stellar envelope ($\sim 10$\,s), this corresponds to a total
accreted mass $\sim 1$\,M$_{\odot}$.  In addition to the accretion
process, an even larger reserve lies in the rotational energy
of the black hole:
\begin{equation}
E_{\mathrm{rot}} = \frac{M_{\mathrm{BH}} c^2}{2} \left\{2 - \left[\left(1 + 
                 \sqrt{1 - a^2}\right)^2 + a^2\right]^{1/2}\right\},
\label{eqn:mrot}
\end{equation}
where $M_{\mathrm{BH}}$ is the black hole mass and $a$ is the 
dimensionless rotation parameter ($a \equiv J c / M_{\mathrm{BH}}^{2} G$).

At least two competing theories exist, however, to explain how the
system can channel this energy to produce the collimated,
relativistic jets we observe from GRBs.  First, the energy may be
extracted from the accretion process via $\nu \bar{\nu}$ annihilation
(e.g, \citealt{w93,mw99,pwf99,rj99,npk01}).  At the hyper-Eddington
rates expected for collapsars, the accretion disk formed around the
rotating black hole will be optically thick and photons are unable to
escape.  The viscous heat is instead balanced by cooling via neutrino
emission.  Annihilation of $\nu \bar{\nu}$ pairs will produce a gas of hot
electron-positron pairs, which, assuming sufficiently low baryon
loading, can then rapidly expand in the low-density regions along the
axis of rotation (a ``fireball'', e.g.,~\citealt{zwm03}).

Alternatively, if the black hole is rotating and the accretion disk is
threaded by sufficiently 
large magnetic fields ($B \approx 10^{15}$\,G for a 3\,M$_{\odot}$ black hole),
energy can be extracted directly from the rotating black hole via the
Blandford-Znajek mechanism (e.g, \citealt{bz77,lwb01}).  The accretion 
disk serves primarily to anchor the
large magnetic field (which would otherwise disperse), and the energy
is emitted as a largely Poynting flux-dominated outflow.

The efficiency of converting the potential energy of the rotating 
black hole plus accretion disk system into a form suitable to launch
a collimated, relativistic outflow has been intensively studied in the last
decade.  In the case of $\nu \bar{\nu}$ annihilation, the overall
efficiency of converting the neutrino luminosity of the cooling disk
($L_{\nu}$) into $e^{+}$--$e^{-}$ pairs ($L_{\nu \bar{\nu}}$) depends sensitively
on the mass accretion rate, the disk geometry, and (perhaps) the effects
of general relativity on neutrino physics (e.g., \citealt{pwf99,mw99,rj99}).
But the majority of recent simulations suggest at most a modest efficiency 
($\eta_{\nu \bar{\nu}} \equiv L_{\nu \bar{\nu}} / L_{\nu} \lesssim 0.5$\%) for
the energy available to launch a relativistic jet (e.g., 
\citealt{rrd03,baj+07,hkt10}).  For typical disk luminosities of
$L_{\nu} \approx 10^{51}$--$10^{52}$\,erg\,s$^{-1}$, this would require an 
extremely long-lived phase of continuous accretion to explain 
hyper-energetic GRBs.

Jets powered by MHD processes, on the 
other hand, can more easily accommodate large energy releases.  To begin
with, by extracting energy directly from the rotation of the black hole,
the Blandford-Znajek process begins with a significantly larger
energy reservoir: $E_{\mathrm{rot}} \approx 8 \times 10^{53}$\,erg for a 
rapidly spinning ($a = 0.9$) black hole of mass 3\,M$_{\odot}$ 
(Eqn.~\ref{eqn:mrot}).  Both analytic models (e.g., \citealt{lwb01})
and numerical simulations (e.g., \citealt{m05}) suggest that as much
as 5--10\% of this rotational energy can be made available to power
a collimated outflow, easily within the requirements of
hyper-energetic events.  We caution, however, that the jet efficiency
is a strong function of the black hole spin, and slowly spinning
black holes have significantly lower efficiencies.

Setting energetics considerations aside for the moment, there still 
remains the question of whether either the $\nu \bar{\nu}$
annihilation mechanism or MHD processes are capable of producing
jets with the Lorentz factors and degree of collimation inferred from
GRB observations.  Here our sample of \fermi-LAT events, with their
extreme initial Lorentz factors, offers a distinct advantage over 
previous GRB studies. Combining the Lorentz-factor 
limits for the most relativistic GRBs with the inferred jet
opening angles from our broadband afterglow models can provide
critical diagnostics of the jet acceleration mechanism.

The results from MHD simulations of jet acceleration appear to depend
sensitively on the nature of the medium into which the jet propagates.
\citet{tnm09} have recently conducted fully relativistic MHD
simulations in which a mildly magnetized (magnetization parameter
$\sigma \lesssim 1$) jet is initially confined by the pressure of a
stellar envelope out to some radius, and then allowed to propagate
freely in all directions (see also \citealt{kvk09}).  Both authors
find that the jet accelerates rapidly in this transition region
(``rarefaction'' acceleration) and can reach Lorentz factors $\Gamma_{0}$
of a few hundred, yet still remains highly collimated
($\theta \lesssim 5^{\circ}$) even after leaving the region of
confinement.  Furthermore, the deconfinement radii required to produce
these outflows agree well with the expected value for Wolf-Rayet stars
($r \approx 10^{9}$--$10^{11}$\,cm).

Based on both results from these simulations and analytical
arguments, \citet{tnm09} demonstrate a relationship
between the initial Lorentz factor and jet opening angle: 
$\Gamma_{0} \theta \approx 10$--30.  This result is able to nicely
reproduce the typical observed Lorentz factors \citep{ls01}
and opening angles \citep{zkk06} of previous GRB samples.
However, for the events studied here, both \grbc\ 
($\Gamma_{0} \theta \approx 70$) and \grbd\ ($\Gamma_{0} \theta
\approx 90$) appear inconsistent with this result.

One way to circumvent this requirement is if the jets have 
extremely high magnetization parameters ($\sigma \gg
1$; \citealt{tmn09b}), as such Poynting flux-dominated
jets can be accelerated to extreme Lorentz factors over 
relatively large angles.  However, it is unclear how such
outflows can convert sufficient electromagnetic energy
to accelerate electrons and produce the observed
prompt gamma-ray emission.  Other particle-acceleration
mechanisms besides MHD shocks may be required in 
this case (e.g., \citealt{b09}).  Alternatively, the gamma-ray
emission may be patchy (e.g., \citealt{kp00}) or the jet may be
structured (see, e.g., \citealt{g07} and references therein), so that 
we are measuring only the extrema of $\Gamma_{0}$ and not
the true bulk of the relativistic flow carrying most of the
energy.

%%%%%%%%%%%%%%%%%%%%%%%%%%%%%%%%%%%%%%%%%%%%%%%%%%%%%%%%%%%%%%%%%%%%%%
\subsection{Comparison with Other Work}
\label{sec:comp}
%%%%%%%%%%%%%%%%%%%%%%%%%%%%%%%%%%%%%%%%%%%%%%%%%%%%%%%%%%%%%%%%%%%%%%
In this section, we attempt to place this work in context, both by comparing 
our results with those of other authors who have studied these same events,
and by highlighting additional differences between our LAT sample and
GRBs detected by satellites at lower energies.

\citet{mkr+10} present optical and NIR observations of three events from this
work (GRBs\,090323, 090328, and 090902B), taken primarily with the 
GROND instrument \citep{gbc+08}.  In the case of \grba, these authors find
an optical spectral index of $\alpha_{O} = 1.90 \pm 0.01$, consistent with
the value derived here, and our measurements of the host-galaxy flux agree
nicely.  We derive a slightly steeper optical spectral index, but our results
are consistent at the 2.5$\sigma$ level.  Most importantly, \citet{mkr+10}
infer from the steep optical decay that a jet break occurred \textit{before}
the first optical observations began ($t_{\mathrm{j}} \lesssim 1$\,day), 
resulting
in a narrow beaming angle ($\theta \lesssim 2^{\circ}$) and a correspondingly
small collimation-corrected prompt energy release [$E_{\gamma} \lesssim
3 (1) \times 10^{51}$\,erg for a constant-density (wind-like) circumburst
medium].  We consider this possibility unlikely, however, as it is
difficult to explain both the flat radio light curve and the more slowly
fading X-ray afterglow ($\alpha_{X} \approx 1.5$) through post jet-break
evolution (see also \S~\ref{sec:limit}).  

Our results also differ from those of \citet{mkr+10} regarding the jet break
time of \grbb.  These authors argue that the steep optical decay index
($\alpha_{o} \approx 2.3$) derived at early times requires a jet break 
before the commencement of observations ($t_{\mathrm{j}} \lesssim 1.5$\,days).
The data presented in our work are not sufficient to uniquely determine the 
optical temporal decay index (particularly given the possible contribution
from an underlying host galaxy).  However, it is again difficult to 
reconcile the shallower X-ray decay ($\alpha_{X} \approx 1.65$) and 
in particular the rising radio light curve at this time with models of 
post-jet break evolution (see also \S~\ref{sec:limit}).

We find very similar results to those of \citet{mkr+10} for the 
temporal and spectral indices, inferred circumburst medium, and
jet break time for \grbc.  In particular, the late-time ($t \approx 23$\,days)
optical observation with the VLT provides strong confirmation that
the radio decline at $t \gtrsim 6$\,days is due to a jet break.  
However, our results diverge when translating the jet break time 
into an opening angle.  Because of the low density we derive for
the circumburst medium of \grbc, a jet break time of $t_{\mathrm{j}}
\approx 6$\,days corresponds to a relatively narrow opening angle.
The resulting constraints on $E_{\gamma}$ are roughly a factor
of 4 less (note that part of this difference is also due to the wider
bandpass \citealt{mkr+10} use to calculate $E_{\gamma,\mathrm{iso}}$).

\citet{psp+10} also present optical and X-ray observations of 
\grbc, most of which were included in our broadband modeling
here.  It is not surprising, then, that we derive similar values for
the optical temporal and spectral indices (the same also holds
for the X-ray bandpass).  The afterglow model derived by these 
authors is broadly similar to ours, but we favor a somewhat
steeper electron index ($p = 2.2$ vs.~$1.8$).  Most importantly,
\citet{psp+10} derive a limit on the opening angle of $\theta
> 6^{\circ}$ (based on a lower limit to the jet break time of 
$t_{\mathrm{j}} > 6$\,days).  Though our derived jet break time is 
consistent with the results of these authors, our inferred opening 
angle is a factor of two smaller due to the lower value we have used 
for the circumburst density.  As a result, the limits we derive for the
prompt gamma-ray energy release are a factor of a few less.

\citet{kb09b} (see also \citealt{kb09}) have also presented an 
analysis of the broadband afterglow of \grbc.  Interestingly,
these authors suggest that the delayed high-energy ($E \gtrsim
100$\,MeV) component observed from several LAT events is in
fact due to the same source as the late-time afterglow emission:
synchrotron radiation from accelerated electrons in the circumburst
medium shock-heated by the outgoing blast wave (i.e., 
external shock emission; see also \citealt{ggn+10}).

The afterglow parameters derived by these authors for \grbc\
differ somewhat from ours, largely due to the fact that they assume
the X-ray bandpass at $t \approx 1$\,day falls below the 
synchrotron cooling frequency (i.e., $\nu_{X} < \nu_{c}$).  While
the observed X-ray spectral and temporal indices at $t \approx 
1$\,day are consistent with this picture, we find that a better fit can 
be achieved for both the X-ray \textit{and} optical data if
$\nu_{X} > \nu_{c}$ at $t \approx 1$\,day (see also \citealt{psp+10}).

Using the best-fit late-time afterglow parameters we have 
derived for \grbc\ (Table~\ref{tab:0902mod}), we have attempted
to calculate the magnitude of the afterglow (i.e., external shock)
flux at $\nu = 100$\,MeV and $t = 50$\,s.  We find 
$f_{\nu} \approx 30$\,nJy, roughly a factor of 7 below the observed
value \citep{aaa+09b}.  The discrepancy between our results
and those of \citet{kb09b} is largely due to the lower kinetic
energy we have inferred for the blast wave ($E_{\mathrm{KE,iso}}
\approx 7 \times 10^{53}$, a factor of 18 lower than that used by
\citealt{kb09b}).  This would suggest that the delayed high-energy
component is not due to afterglow emission.  
We caution, however, that our flux calculation
does not incorporate the reduced cross-section for
inverse Compton emission due to the Klein-Nishina effect at
very high energies, and this will affect our flux calculation
to some extent.

%%%%%%%%%%%%%%%%%%%%%%%%%%%%%%%%%%%%%%%%%%%%%%%%%%%%%%%%%%%
\begin{figure}
\epsscale{1.1}
\plotone{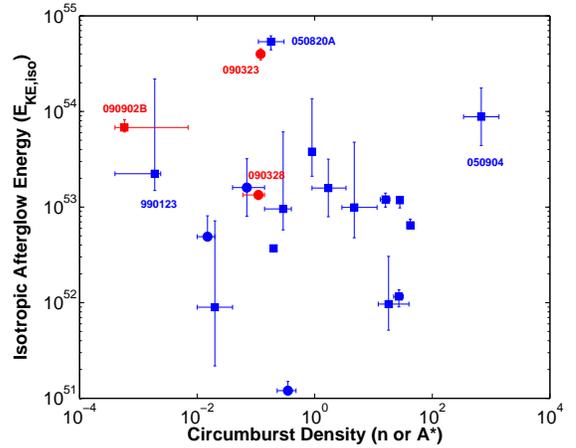}
%\plotone{../Figures/eken.eps}
\caption{Circumburst density and isotropic afterglow energy release from 
cosmological ($z > 0.5$) GRBs. 
Constant-density models are shown as squares, while wind-like models are 
plotted as circles.  The GRBs detected by the \fermi-LAT (red) clearly have 
preferentially larger isotropic energy releases and smaller circumburst 
densities than the rest of the sample.  See the caption of 
Figure~\ref{fig:energy} for references.}
\label{fig:density}
\end{figure}
%%%%%%%%%%%%%%%%%%%%%%%%%%%%%%%%%%%%%%%%%%%%%%%%%%%%%%%%%%%%

A further interesting claim from the work of \citet{kb09b} is
that the inferred value of the magnetic field for \grbc\ (along 
with several other LAT-detected events) is consistent with 
shock compression of a modest  circumstellar 
field ($B \gtrsim 30$\,$\mu$G).  In other words, no dynamo
process is necessary to generate the magnetic field strengths
needed to produce the observed synchrotron afterglow 
emission.  For a constant-density circumburst medium,
the preshock magnetic field is given by
\begin{equation}
B = (2 \pi m_{p} \epsilon_{B} n)^{1/2} c.
\label{eqn:Bism}
\end{equation}
We also find that the product $\epsilon_{B} \times n$ (and hence the 
derived $B$ field) is smaller for LAT events than for other GRBs detected at 
lower energies, primarily due to the lower inferred circumburst densities (see 
below).  However, using our best-fit parameters for \grbc, we find
$B \approx 100$\,$\mu$G, broadly consistent with the results of \citet{pn10}
and \citet{l10}.  While this value is smaller than the preshock
magnetic fields previously inferred for GRBs ($B \approx 10$\,mG
for $\epsilon_{B} = 0.01$, $n = 1$\,cm$^{-3}$), it suggests that
at least some magnetic field amplification may still be required.

Finally, we return to the issue of the relatively low circumburst
densities derived in $\S$~\ref{sec:results}.  Given the strong 
evidence for a connection between long-duration GRBs and
broad-lined SNe Ib/c, it would be natural to expect the massive-star 
progenitors of GRBs to explode in the dense molecular
cloud environments where we observe stars forming in our own
Galaxy.

Past modeling of broadband GRB afterglows, however, has not always
revealed this to be the case.  Of the past GRBs with sufficient
radio observations to estimate the circumburst density, the derived
values span a large range from $1.9 \times 10^{-3}$\,cm$^{-3}$
(GRB\,990123; \citealt{pk02}) to $680$\,cm$^{-3}$ (GRB\,050904;
\citealt{fck+06}).  We plot in Figure~\ref{fig:density} the derived 
circumburst densities for
these previous events (either $n$ or $A_{*}$, depending on the 
circumburst density profile) as a function of isotropic afterglow
kinetic energy, along with three events from our LAT sample (we have
not included \grbd\, as lack of radio coverage makes density
estimates highly degenerate).  It is clear that the events detected
by the LAT have on average larger isotropic energies and smaller densities 
than the previous sample.  

The larger isotropic kinetic energy is not hard to understand, as the LAT is
less sensitive than, for example, the Burst Alert Telescope (BAT; 
\citealt{bbc+05}) onboard \swift\ and should therefore select
brighter events (in terms of the high-energy fluence).  Unless 
$E_{\gamma,\mathrm{iso}}$ and $E_{\mathrm{KE,iso}}$ were anticorrelated, we would
expect LAT events to have larger isotropic blast-wave energies as well.
But the lower densities require an additional explanation.  We speculate
on possible reasons for this difference in \S~\ref{sec:conclusions}.

The large ratio of $E_{\mathrm{KE,iso}} / n$ (or, alternatively 
$E_{\mathrm{KE,iso}} / A_{*}$) we have derived for these events has two
important implications.  First, the opening angle derived for a given
jet break time scales inversely as $E_{\mathrm{KE,iso}}$ and proportionally
to $n / A_{*}$ (\S~\ref{sec:limit}).  Thus, for a given jet break time, the 
large ratios we have inferred for LAT events imply smaller opening angles 
than were one to simply use a canonical value (e.g., $n = 0.1$\,cm$^{-3}$
from \citealt{fks+01}; $n = 10$\,cm$^{-3}$ from \citealt{bfk03}).  In this 
manner, LAT events can have relatively late jet breaks but still be narrowly 
collimated ($\theta \lesssim 10^{\circ}$).

Secondly, a large $E_{\mathrm{KE,iso}} / n$ ratio will act to delay the 
deceleration time of the outflow.  Assuming the ``thin shell''
case, the outgoing relativistic blast wave will decelerate when
\citep{sp99,mvm+07}
\begin{eqnarray}
\label{eqn:gammaISM}
t_{\mathrm{dec}} (\mathrm{ISM}) & = & \left(\frac{3 E_{\mathrm{KE,iso}} (1 + z)^{3}}
                                {32 \pi m_{p} c^{5} n \Gamma_{0}^{8}}\right)^{1/3}, \\
t_{\mathrm{dec}} (\mathrm{Wind}) & = & \frac{E_{\mathrm{KE,iso}} (1 + z)}
                                {8 \pi c^{3} A \Gamma_{0}^{4}}.
\label{eqn:gammawind}
\end{eqnarray}
(Note that we have corrected Eqn.~2 from \citealt{mvm+07} to remove
the erroneous factor of $m_{p}$.)  To some extent, the large ratio of 
$E_{\mathrm{KE,iso}} / n$ will offset the effect of the larger initial Lorentz
factors for \fermi-LAT events.  Using the parameters we have derived 
from the LAT events in our sample, we find $t_{\mathrm{dec}} 
\approx 2$\,s for \grba, $t_{\mathrm{dec}} \approx 3$\,s for \grbb, and 
$t_{\mathrm{dec}} \approx 33$\,s for \grbc. If the late-time GeV emission is 
indeed due to external shock emission, the delay between the MeV and GeV 
photons should correspond roughly to the blast-wave deceleration
time.  This may be problematic for \grbc, for which the observed
delay ($\sim 4$\,s) is significantly less than our inferred 
deceleration time.  We caution, however, that the large systematic
uncertainties in $E_{\mathrm{KE,iso}}$, $n$, and particularly 
$\Gamma_{0}$ (\S~\ref{sec:limit}) may introduce a significant 
error into calculations of the deceleration time.

%%%%%%%%%%%%%%%%%%%%%%%%%%%%%%%%%%%%%%%%%%%%%%%%%%%%%%%%%%%%%%%%%%%%%%
\subsection{Limitations and Future Work}
\label{sec:limit}
%%%%%%%%%%%%%%%%%%%%%%%%%%%%%%%%%%%%%%%%%%%%%%%%%%%%%%%%%%%%%%%%%%%%%%

To better understand our results in \S~\ref{sec:engines}, 
\S~\ref{sec:acceleration}, and
\S~\ref{sec:comp} on GRB energetics and circumburst density, it is
important to consider the robustness of our modeling. The most crucial
factor in determining the true energy release is the collimation
correction, which we therefore examine in more detail here.

For a wind-like circumburst medium, the jet half-opening angle $\theta$ can
be written as (e.g., \citealt{cl00})
\begin{align}
\theta(\mathrm{Wind}) = & 0.10 \left(\frac{1+z}{2}\right)^{-1/4}
    \left(\frac{E_{\mathrm{KE,iso}}}{10^{52}\,\mathrm{erg}}\right)^{-1/4}
    \nonumber \\
    & \left(\frac{A_{*}}{1.0\,\mathrm{cm}^{-1}}\right)^{1/4}
    \left(\frac{t_{\mathrm{j}}}{1\,\mathrm{d}}\right)^{1/4}\,\,\,[\mathrm{rad}].
\label{eqn:thetawind}
\end{align}
This result assumes only that the shock is ultrarelativistic and undergoes a
self-similar evolution \citep{bm76}, and is therefore relatively robust.

Only in the case of \grbd\ (and also for \grbc\ if we include the late-time 
VLT observation from \citealt{mkr+10}) do we find clear evidence for a jet 
break in multiple bandpasses.  For \grba, the lack of a steepening in the
X-ray and optical light curves apparently limits $t_{\mathrm{j}} \gtrsim 
10$\,days.  If we simply ignore our modeling results and assume that the shock 
energy is comparable to the prompt gamma-ray energy ($E_{\mathrm{KE,iso}}
\approx 4 \times 10^{54}$\,erg) and that the progenitor wind speed and
mass-loss rate are comparable to those observed from Galactic Wolf-Rayet 
stars ($A_{*} \approx 1$\,g\,cm$^{-1}$), then we find a lower limit on the 
opening angle of $\theta$(Wind) $\gtrsim 3^{\circ}$.  The resulting limit on 
the prompt energy release is therefore $E_{\gamma} \gtrsim 5 \times
10^{51}$\,erg, similar to what we derived from our broadband models.

If we were instead to assume the GRB exploded in a constant-density
medium, the opening angle then becomes (e.g., \citealt{sph99})
\begin{align}
\theta(\mathrm{ISM}) = & 0.12 \left(\frac{1+z}{2}\right)^{-3/8}
   \left(\frac{E_{\mathrm{KE,iso}}}{10^{52}\,\mathrm{erg}}\right)^{-1/8}
   \nonumber \\
   & \left(\frac{n}{1\,\mathrm{cm}^{-3}}\right)^{1/8}
   \left(\frac{t_{\mathrm{j}}}{1 \mathrm{d}}\right)^{3/8}\,\,\,[\mathrm{rad}].
\label{eqn:thetaISM}
\end{align}
Under similar assumptions as above, the limit on the opening angle would be
$\theta$(ISM) $\gtrsim 6^{\circ}$, corresponding to a prompt energy release of
$E_{\gamma} \gtrsim 2 \times 10^{52}$\,erg.  Clearly a constant-density
circumburst medium only makes the energy requirements more strict.

An alternative possibility for \grba\ (\citealt{mkr+10}; also suggested 
by \citealt{sdp+07b} to explain
the afterglow of GRB\,061007) is a jet break before the beginning
X-ray and optical observations ($t \lesssim 1$\,day for \grba).  Much
like GRB\,061007, however, the implied post-break decay index in the
X-ray ($\alpha_{X} \approx 1.5$) is
too shallow to accommodate a typical electron spectral index $p$ (for
$\nu > \nu_{m}$, $f_{\nu} \propto t^{-p}$ post jet-break;
\citealt{sph99}).  The observed behavior would therefore require
late-time energy input to flatten the X-ray and optical decay.
Furthermore, the relatively flat radio light curve for $t \lesssim
20$\,days would require that the synchrotron self-absorption frequency
fell above the radio bandpass during this phase; otherwise the radio
bandpass would be required to decay, either as $t^{-1/3}$ ($\nu <
\nu_{m}$) or $t^{-p}$ ($\nu > \nu_{m}$)\footnote{We note that the above
results are independent of circumburst medium, as the expansion of
the outflow is predominantly lateral after the jet break
\citep{cl00}.}.  We consider this somewhat contrived picture
unlikely in the case of \grba, particularly given the reasonable
quality of our fits using only standard afterglow theory.

A similar analysis of \grbb\ suggests that the jet-break time cannot
occur any earlier than $t \approx 4$\,days, based again on the relatively
steady decay seen in the X-ray light curve and, in particular, on the
rising radio emission at early times.  The
corresponding limits on the opening angles are therefore
$\theta$(Wind) $\gtrsim 8^{\circ}$, $\theta$(ISM) $\gtrsim 7^{\circ}$.
Both values result in a prompt energy release $E_{\gamma} \gtrsim
10^{51}$\,erg, broadly consistent with our modeling results.  

In the end, all of these estimates of energy, density, and jet geometry
rest on the relativistic synchrotron model. While the standard
afterglow model has undergone continuous improvements and it has been
well tested \citep{mr97,spn98,sph99,wg99,cl00,pk00,yhs+03,woo+07}, it
still must make several simplifying assumptions about the shock
dynamics, magnetic field generation, particle acceleration, energy
injection, and the circumburst density structure.  Of particular concern
are the recent results of relativistic simulations by \citet{zm09}, which 
suggest that viewing angle can dramatically alter the observed afterglow
emission at early times.

If we are to verify
this class of hyper-energetic bursts, we need independent estimates of
the relativistic energy content. \citet{wkf98} first suggested that
late-time (radio) calorimetry could be used for this purpose in order
to sidestep early-time complications such as the outflow
geometry, the density structure, and ongoing activity from the central
engine. Recent relativistic hydrodynamic simulations show that the
earlier analytic models which used spherical geometry and Sedov-Taylor
dynamics to explain the late-time behavior were approximately correct 
and produce a robust estimate of
the total kinetic energy \citep{zm09}. In the past, the method has been
limited to only a small number of radio-bright events (e.g.,
\citealt{vkr+08} and references therein).  But it is somewhat encouraging
that both early- and late-time estimates for the most energetic
pre-\fermi\ event (GRB\,970508; $E_{\mathrm{KE}} \approx 10^{52}$\,erg)
are in close agreement \citep{yhs+03,bkf04}.  With the advent of radio
facilities having greatly increased sensitivity such as LOFAR \citep{v05b}
and the EVLA \citep{pnj+09},
it should soon be possible to verify candidate hyper-energetic
events by searching for long-lived ($t \gtrsim 100$\,days) radio
afterglows.

Finally, we add a further note of caution that the derivation of the 
initial Lorentz factor $\Gamma_{0}$ is also subject to significant
systematic uncertainties (e.g., \citealt{bdd09}).  In order to calculate 
the optical depth to pair production, it is necessary to determine the
effective blast-wave radius as well as the instantaneous spectral
parameters.  The former is typically inferred from the variability
time scale, while the latter is usually averaged over the entire duration
of the prompt emission, and it is not always clear how large an 
effect this will have on the calculation of $\Gamma_{0}$.  The
calculations are further complicated if the spectrum is not a
pure power law, as would be the case for thermal emission arising
from the jet photosphere \citep{raz+10}.  Here we have attempted to 
calculate the initial Lorentz
factors in a consistent manner, and it is not unreasonable to 
believe that the relative ordering of $\Gamma_{0}$ is robust.  However,
small changes in $\Gamma_{0}$ can have a large effect on some of the
parameters derived here (e.g., $t_{\mathrm{dec}}$; 
Eqns.~\ref{eqn:gammaISM} and \ref{eqn:gammawind}).  Future 
early afterglow observations of \fermi-LAT GRBs to directly 
measure the deceleration time and hence constrain 
$\Gamma_{0}$ (e.g., \citealt{mvm+07}) could provide valuable
insight in this area.

%%%%%%%%%%%%%%%%%%%%%%%%%%%%%%%%%%%%%%%%%%%%%%%%%%%%%%%%%%%%%%%%%%%%%%%%
\section{Conclusions}
\label{sec:conclusions}
%%%%%%%%%%%%%%%%%%%%%%%%%%%%%%%%%%%%%%%%%%%%%%%%%%%%%%%%%%%%%%%%%%%%%%%%
We have undertaken extensive broadband continuum (radio, optical, and
X-ray) and spectroscopic observations of four long-duration GRBs
(GRBs\,090323, 090328, 090902B, and 090926A) detected by the LAT 
instrument on the \fermi\ satellite at GeV energies.
This work was motivated by the realization that \fermi\ is especially
sensitive to GRBs with large isotropic energy release, and hence
provides an interesting sample of events to test GRB
central-engine models and their relativistic outflows. Our afterglow
models constrain the jet break times and the density of the
circumburst medium, from which we derive the collimation angle and
hence beaming-corrected energy release for each event. We find three
GRBs with a total relativistic content about an order of magnitude in
excess of the canonical 10$^{51}$ erg, with \grbd\ almost certainly
in excess of $10^{52}$\,erg. This analysis provides support
for our earlier claim of a class of hyper-energetic GRBs. The
discovery of more GRBs with total energy release $\gtrsim
10^{52}$\,erg is troubling for central engines in which the energy to
drive the jet is derived either from a rotating magnetar or
collapsars powered by neutrino annihilation.  For this reason,
we are led to believe that, at least for hyper-energetic GRBs, the
massive star progenitor collapses directly to a black hole and the
rotational energy of this system is extracted via the Blandford-Znajek
process.

Although we find relatively narrow opening angles for all four events
($\theta \lesssim 10^{\circ}$), the extreme initial Lorentz factors
inferred for these LAT events imply that the product $\theta \Gamma_{0}$
can be a factor of 5--10 larger than estimates of previous GRBs 
detected at MeV energies.  These values are inconsistent with
recent simulations of low-magnetization MHD jets, suggesting that the
outflow may be at least initially Poynting-flux dominated.  If this
is indeed the case, it is unclear how the initial kinetic energy of
the outflow is converted to prompt gamma-ray emission.

Interestingly, for the three events having sufficient radio coverage to 
derive a circumburst density, we find anomalously large values
of $E_{\mathrm{KE,iso}} / n$ (or, for a wind-like medium, 
$E_{\mathrm{KE,iso}} / A_{*}$).  While the large $E_{\mathrm{KE,iso}}$
values are simple to understand, the low circumburst densities
require a more complex explanation.  

One possibility is that the progenitor stars of LAT GRBs are
somehow different from the progenitors of most previous GRBs
detected at MeV or keV energies.  It is currently thought that GRB
progenitors are distinguished from the progenitors of ordinary
SNe Ib/c by their low metallicities (e.g., \citealt{mw99,wh06,mkk+08}): 
the lower mass-loss rates allow the progenitors of GRBs to keep
more angular momentum.  The increased rotation evacuates a 
cavity through which a relativistic jet can propagate.  If LAT
events have larger initial Lorentz factors, it may be that they
come from lower metallicity progenitors with minimal pre-explosion
mass loss.  Observations of the host galaxies of these events,
both through absorption and emission spectroscopy, may help
shed light on this matter.

It is also possible that the low density preference is the result of
other, more subtle, selection effects.  In particular, if the GeV 
emission arises in the external shock, the \fermi-LAT could be
biased towards events in low-density environments.  If the 
circumburst density is too high, the blastwave will decelerate
at small radii (Equations~\ref{eqn:gammaISM} and 
\ref{eqn:gammawind}), where the outflow may be opaque to
GeV photons.  More observations of LAT GRBs, particularly at
very early times, would help to investigate this hypothesis.

We end by emphasizing the importance of afterglow observations of
high-$E_{\gamma,\mathrm{iso}}$ events in the \fermi\ era to provide
further confirmation of this picture. Such GRBs are either highly
collimated outflows ($\theta \lesssim 2^{\circ}$) with a typical
energy release, or truly hyper-energetic events; both represent
extreme tests of jet collimation and central-engine models,
respectively. Current efforts suffer from delays in LAT localizations
and limited ground-based afterglow follow-up efforts. The latter can be
improved by focusing rare follow-up resources on \fermi-LAT GRBs;
as \citet{nfp09} and \citet{mkr+10} have shown, these events have brighter
X-ray and optical afterglows on average, and are therefore accessible
even for moderate-aperture optical facilities.  Targeting these bright
afterglows will make it easier to measure the jet breaks, which have
proven almost impossible to obtain in the \swift\ era. Finally, we note
that one testable consequence of hyper-energetic GRBs is long-lived
afterglow emission ($\gtrsim 1$\,yr). If the shock microphysics and the
circumburst density do not undergo a drastic evolution, it should be
possible to detect these afterglows with the upcoming generation of
radio facilities and carry out calorimetry measurements. 

%%%%%%%%%%%%%%%%%%%%%%%%%%%%%%%%%%%%%%%%%%%%%%%%%%%%%%%%%%%%%%%%%%%%%%%%%%%
\acknowledgements
S.B.C.~and A.V.F.~wish to acknowledge generous support from Gary and
Cynthia Bengier, the Richard and Rhoda Goldman Fund, National Aeronautics and
Space Administration (NASA)/\swift\ 
grants NNX09AL08G and NNX10AI21G, and 
National Science Foundation (NSF) grants AST--0607485 and AST--0908886.  
B.E.C.~gratefully acknowledges support from an NSF Astronomy \& Astrophysics 
Postdoctoral Fellowship (AST-0802333).  N.R.B.~is supported through the 
Einstein Fellowship Program (NASA Cooperative Agreement NNG06DO90A).
J.~S.~B.~and his group were partially supported by NASA/\swift\ Guest
Investigator grant NNX09AQ66G and a grant from DOE SciDAC.

P60 operations are funded in part by NASA through the
{\it Swift} Guest Investigator Program (grant number NNG06GH61G).
Based in part on observations obtained at the Gemini Observatory
(Programs GS-2009A-Q-23, GS-2009B-Q-5, GN-2009A-Q-26, and
GN-2009B-Q-28), which is operated by the Association of Universities for
Research in Astronomy, Inc., under a cooperative agreement with the
NSF on behalf of the Gemini partnership: the National Science
Foundation (US), the Particle Physics and Astronomy Research Council
(UK), the National Research Council (Canada), CONICYT (Chile), the
Australian Research Council (Australia), CNPq (Brazil) and CONICET
(Argentina).  We wish to thank the entire staff at Gemini for assistance
with these observations.  SMARTS is supported by NSF grant AST--0707627.

%%%%%%%%%%%%%%%%%%%%%%%%%%%%%%%%%%%%%%%%%%%%%%%%%%%%%%%%%%%%%%%%%%%%%%%%%%%
{\it Facilities:} \facility{Fermi (LAT, GBM)}, \facility{VLA}, 
  \facility{PO:1.5m}, \facility{Gemini:South (GMOS)}, 
  \facility{Gemini:North (GMOS)}, \facility{CTIO:2MASS (ANDICAM)},
  \facility{Swift (XRT, UVOT)}

%%%%%%%%%%%%%%%%%%%%%%%%%%%%%%%%%%%%%%%%%%%%%%%%%%%%%%%%%%%%%%%%%%%%%%%%%%%

%%%%%%%%%%%%%%%%%%%%%%%%%%%%%%%%%%%%%%%%%%%%%%%%%%%%%%%%%%%%%%%%%%%%%%%%%%%%
 
\clearpage

%%%%%%%%%%%%%%%%%%%%%%%%%%%%%%%%%%%%%%%%%%%%%%%%%%%%%%%%%%%%%%%%%%%%%%%%%%%
% Tables
%%%%%%%%%%%%%%%%%%%%%%%%%%%%%%%%%%%%%%%%%%%%%%%%%%%%%%%%%%%%%%%%%%%%%%%%%%%

%%%%%%%%%%%%%%%%%%%%%%%%%%%%%%%%%%%%%%%%%%%%%%%%%%%%%%%%%%%%
\begin{deluxetable}{lcccccr}
\tabletypesize{\scriptsize}
\tablecaption{Optical/NIR Observations of \grba}
\tablewidth{0pt}
\tablehead{
\colhead{Date\tablenotemark{a}} & \colhead{Time Since Burst\tablenotemark{b}} & \colhead{Telescope/Instrument} & \colhead{Filter} & \colhead{Exposure Time} & \colhead{Magnitude\tablenotemark{c}} & \colhead{Reference\tablenotemark{d}}\\
\colhead{(UT)} & \colhead{(days)} & & & \colhead{(s)} & &
}
\startdata
% GCN9026
2009 Mar 24.12 & 1.12 & GROND & $g^{\prime}$ & 480.0 & $21.64 \pm 0.07$ & 1 \\ 
2009 Mar 24.12 & 1.12 & GROND & $r^{\prime}$ & 480.0 & $20.03 \pm 0.03$ & 1 \\
2009 Mar 24.12 & 1.12 & GROND & $i^{\prime}$ & 480.0 & $19.64 \pm 0.02$ & 1 \\
2009 Mar 24.12 & 1.12 & GROND & $z^{\prime}$ & 480.0 & $19.39 \pm 0.02$ & 1 \\
2009 Mar 24.12 & 1.12 & GROND & $J$ & 480.0 & $19.24 \pm 0.02$ & 1 \\
2009 Mar 24.12 & 1.12 & GROND & $H$ & 480.0 & $18.86 \pm 0.02$ & 1 \\
2009 Mar 24.12 & 1.12 & GROND & $K_{s}$ & 480.0 & $18.58 \pm 0.03$ & 1 \\
2009 Mar 24.19 & 1.20 & P60 & $r^{\prime}$ & 1800.0 & $20.50 \pm 0.07$ & * \\
2009 Mar 24.20 & 1.21 & P60 & $i^{\prime}$ & 1800.0 & $19.95 \pm 0.04$ & * \\
2009 Mar 24.23 & 1.23 & Gemini-S/GMOS & $i^{\prime}$ & 90.0 & $20.04 \pm 0.03$ & * \\
2009 Mar 24.32 & 1.32 & P60 & $r^{\prime}$ & 900.0 & $20.46 \pm 0.04$ & * \\
2009 Mar 24.33 & 1.33 & P60 & $i^{\prime}$ & 900.0 & $20.21 \pm 0.04$ & * \\
2009 Mar 24.34 & 1.35 & P60 & $r^{\prime}$ & 900.0 & $20.56 \pm 0.04$ & * \\
2009 Mar 24.36 & 1.36 & P60 & $i^{\prime}$ & 900.0 & $20.20 \pm 0.04$ & * \\
2009 Mar 24.42 & 1.42 & P60 & $r^{\prime}$ & 900.0 & $20.65 \pm 0.05$ & * \\
2009 Mar 24.43 & 1.43 & P60 & $i^{\prime}$ & 900.0 & $20.36 \pm 0.05$ & * \\
2009 Mar 24.44 & 1.44 & P60 & $r^{\prime}$ & 900.0 & $20.60 \pm 0.04$ & * \\
2009 Mar 24.45 & 1.46 & P60 & $i^{\prime}$ & 900.0 & $20.34 \pm 0.06$ & * \\
% GCN 9034
2009 Mar 24.65 & 1.65 & Xinglong/TNT & $R$ & 6000.0 & 20.71 & 2\\
% GCN 9033
2009 Mar 24.88 & 1.88 & TLS Tautenberg & $R$ & 4800.0 & $21.04 \pm 0.04$ & 3 \\
% GCN 9037
2009 Mar 25.06 & 2.06 & RTT150 & $R$ & 3600.0 & $21.01 \pm 0.04$ & 4 \\
2009 Mar 25.21 & 2.21 & P60 & $r^{\prime}$ & 720.0 & $21.51 \pm 0.10$ & * \\
2009 Mar 25.22 & 2.22 & P60 & $i^{\prime}$ & 900.0 & $20.96 \pm 0.09$ & * \\
2009 Mar 25.23 & 2.24 & P60 & $r^{\prime}$ & 900.0 & $21.54 \pm 0.11$ & * \\
2009 Mar 25.24 & 2.25 & P60 & $i^{\prime}$ & 900.0 & $21.05 \pm 0.09$ & * \\
2009 Mar 25.26 & 2.26 & P60 & $r^{\prime}$ & 720.0 & $21.76 \pm 0.09$ & * \\ 
2009 Mar 25.27 & 2.27 & P60 & $i^{\prime}$ & 720.0 & $21.24 \pm 0.08$ & * \\
2009 Mar 25.28 & 2.28 & P60 & $r^{\prime}$ & 720.0 & $21.72 \pm 0.08$ & * \\
2009 Mar 25.29 & 2.29 & P60 & $i^{\prime}$ & 900.0 & $21.33 \pm 0.08$ & * \\
% GCN 9036
2009 Mar 25.30 & 2.30 & Nickel & $R$ & 3300.0 & $21.46 \pm 0.2$ & 5 \\
2009 Mar 25.30 & 2.31 & P60 & $r^{\prime}$ & 720.0 & $21.61 \pm 0.07$ & * \\
2009 Mar 25.31 & 2.32 & P60 & $i^{\prime}$ & 720.0 & $21.25 \pm 0.07$ & * \\
% GCN 9039
2009 Mar 25.50 & 2.50 & Faulkes South & $i^{\prime}$ & 1800.0 & $21.3 \pm 0.2$ & 6 \\
2009 Mar 25.53 & 2.53 & Faulkes South & $R$ & 1800.0 & $21.7 \pm 0.1$ & 6 \\
2009 Mar 26.23 & 3.25 & P60 & $r^{\prime}$ & 3240.0 & $22.29 \pm 0.14$ & * \\
2009 Mar 26.24 & 3.26 & P60 & $i^{\prime}$ & 2880.0 & $22.38 \pm 0.19$ & * \\
2009 Mar 27.31 & 4.31 & P60 & $i^{\prime}$ & 5400.0 & $> 22.60$ & * \\
% GCN 9042
2009 Mar 27.36 & 4.36 & Nickel & $R$ & 15600.0 & $22.79\pm 0.18$ & 7 \\
% GCN 9041
2009 Mar 28.11 & 5.10 & TLS Tautenberg & $R$ & 6600.0 & $22.83 \pm 0.20$ & 8 \\
2009 Mar 28.25 & 5.28 & P60 & $i^{\prime}$ & 4500.0 & $22.92 \pm 0.14$ & * \\
% GCN 9063
2009 Mar 28.89 & 5.89 & TLS Tautenberg & $R$ & 2400.0 & $23.29 \pm 0.50$ & 9 \\
% GCN 9324
2009 Mar 29.00 & 5.99 & Shajn & $R$ & 5460.0 & $22.86 \pm 0.1$ & 10 \\
% GCN 9051
2009 Mar 29.32 & 6.33 & NOT & $R$ & 1800.0 & $22.96 \pm 0.06$ & 11\\
% GCN 9063
2009 Mar 31.91 & 8.90 & TLS Tautenberg & $R$ & 7200.0 & $23.80 \pm 0.36$ & 9 \\
2009 Apr 6.40 & 14.41 & Gemini-N/GMOS & $r^{\prime}$ & 1200.0 & $23.94 \pm 0.10$ & * \\
2009 Jul 29.96 & 129.97 & Gemini-S/GMOS & $r^{\prime}$ & 1800.0 & $24.95 \pm 0.15$ & * \\
\enddata
\tablenotetext{a}{UT at beginning of exposure.}
\tablenotetext{b}{Time from midpoint of exposure to \fermi-GBM trigger.}
\tablenotetext{c}{Reported magnitudes have not been corrected for Galactic 
   extinction ($E(B-V) = 0.025$\,mag; \citealt{sfd98}).  Observations in the 
   $R$ band are referenced to Vega, while all other filters are reported on 
   the AB magnitude system \citep{og83}.}
\tablenotetext{d}{* -- This work; 1 -- \citet{GCN.9026}; 2 -- \citet{GCN.9034}; 
   3 -- \citet{GCN.9033}; 4 -- \citet{GCN.9037}; 5 -- \citet{GCN.9036}; 6 -- 
   \citet{GCN.9039}; 7 -- \citet{GCN.9042}; 8 -- \citet{GCN.9041}; 9 --
   \citet{GCN.9063}; 10 -- \citet{GCN.9324}; 11 -- \citet{GCN.9051}.}
\label{tab:0323opt}
\end{deluxetable}
%%%%%%%%%%%%%%%%%%%%%%%%%%%%%%%%%%%%%%%%%%%%%%%%%%%%%%%%%%%%

\clearpage

%%%%%%%%%%%%%%%%%%%%%%%%%%%%%%%%%%%%%%%%%%%%%%%%%%%%%%%%%%%%
\begin{deluxetable}{lcccccr}
\tabletypesize{\scriptsize}
\tablecaption{Optical/NIR Observations of \grbb}
\tablewidth{0pt}
\tablehead{
\colhead{Date\tablenotemark{a}} & \colhead{Time Since Burst\tablenotemark{b}} & \colhead{Telescope/Instrument} & \colhead{Filter} & \colhead{Exposure Time} & \colhead{Magnitude\tablenotemark{c}} & \colhead{Reference\tablenotemark{d}}\\
\colhead{(UT)} & \colhead{(days)} & & & \colhead{(s)} & &
}
\startdata
2009 Mar 29.06 & 0.67 & UVOT & $U$ & 793.1 & $19.38 \pm 0.10$ & * \\
2009 Mar 29.13 & 0.73 & UVOT & $U$ & 462.1 & $19.58 \pm 0.16$ & * \\
2009 Mar 29.98 & 1.58 & Gemini-S/GMOS & $i^{\prime}$ & 150.0 & 
  $19.98 \pm 0.10$ & * \\
2009 Mar 29.98 & 1.60 & GROND & $g^{\prime}$ & 480.0 & $20.97 \pm 0.05$ & 1 \\
2009 Mar 29.98 & 1.60 & GROND & $r^{\prime}$ & 480.0 & $20.23 \pm 0.03$ & 1 \\
2009 Mar 29.98 & 1.60 & GROND & $i^{\prime}$ & 480.0 & $19.89 \pm 0.04$ & 1 \\
2009 Mar 29.98 & 1.60 & GROND & $z^{\prime}$ & 480.0 & $19.54 \pm 0.03$ & 1 \\
2009 Mar 29.98 & 1.60 & GROND & $J$ & 480.0 & $19.54 \pm 0.06$ & 1 \\
2009 Mar 29.98 & 1.60 & GROND & $H$ & 480.0 & $19.02 \pm 0.06$ & 1 \\
2009 Mar 29.98 & 1.60 & GROND & $K$ & 480.0 & $18.52 \pm 0.08$ & 1 \\
2009 Mar 30.06 & 1.69 & UVOT & $U$ & 4111.2 & $20.80 \pm 0.11$ & * \\
2009 Mar 31.42 & 3.04 & UVOT & $U$ & 4162.2 & $21.96 \pm 0.20$ & * \\
2009 Apr 3.79 & 6.40 & UVOT & $U$ & 2094.3 & $21.52 \pm 0.24$ & * \\
2009 Apr 4.79 & 7.40 & UVOT & $U$ & 2507.4 & $21.69 \pm 0.23$ & * \\
2009 Apr 9.68 & 12.33 & UVOT & $U$ & 9493.6 & $22.67 \pm 0.24$ & * \\
\enddata
\tablenotetext{a}{UT at beginning of exposure.}
\tablenotetext{b}{Time from midpoint of exposure to \fermi-GBM trigger.}
\tablenotetext{c}{Reported magnitudes have not been corrected for Galactic 
   extinction ($E(B-V) = 0.057$\,mag; \citealt{sfd98}).  Observations in the 
   $U$ band are referenced to Vega, while all other filters are reported on 
   the AB magnitude system \citep{og83}.}
\tablenotetext{d}{* -- This work; 1 -- \citet{GCN.9054}.}
\label{tab:0328opt}
\end{deluxetable}
%%%%%%%%%%%%%%%%%%%%%%%%%%%%%%%%%%%%%%%%%%%%%%%%%%%%%%%%%%%%%%%%%%%%%%%%%%%%%

%%%%%%%%%%%%%%%%%%%%%%%%%%%%%%%%%%%%%%%%%%%%%%%%%%%%%%%%%%%%%%%%%%%%%%%%%%%%%
\begin{deluxetable}{lcccccr}
\tabletypesize{\scriptsize}
\tablecaption{Optical Observations of \grbc}
\tablewidth{0pt}
\tablehead{
  \colhead{Date\tablenotemark{a}} & 
  \colhead{Time Since Burst\tablenotemark{b}} & 
  \colhead{Telescope/Instrument} & \colhead{Filter} & 
  \colhead{Exposure Time} & \colhead{Magnitude\tablenotemark{c}} & 
  \colhead{Reference\tablenotemark{d}} \\ \colhead{(UT)} & \colhead{(days)} 
  & & & \colhead{(s)} & &
}
\startdata
2009 Sep 2.51 & 0.06 & ROTSE-IIIa & $R$ & 890.0 & $16.4 \pm 0.5$ & 1 \\
2009 Sep 2.72 & 0.27 & ROTSE-IIId & $R$ & 846.0 & $> 18.7$ & 1 \\
2009 Sep 2.98 & 0.53 & UVOT & $U$ & 1074.7 & $20.31 \pm 0.15$ & * \\
2009 Sep 3.08 & 0.64 & UVOT & $U$ & 3612.9 & $20.54 \pm 0.09$ & * \\
2009 Sep 3.19 & 0.74 & Nickel & $R$ & 3300.0 & $20.60 \pm 0.10$ & 1 \\
2009 Sep 3.24 & 0.79 & Gemini-N/GMOS & $r^{\prime}$ & 180.0 & $20.68 \pm 0.15$
  & * \\
2009 Sep 3.24 & 0.82 & UVOT & $U$ & 8756.0 & $20.92 \pm 0.08$ & * \\
2009 Sep 3.33 & 0.88 & Liverpool & $R$ & 1800.0 & $21.04 \pm 0.11$ & 1 \\
2009 Sep 3.74 & 1.30 & UVOT & $U$ & 3852.8 & $22.06 \pm 0.26$ & * \\
2009 Sep 3.87 & 1.42 & WHT-LIRIS & $J$ & 1800.0 & $19.99 \pm 0.15$ & 1 \\
2009 Sep 3.89 & 1.44 & Liverpool & $R$ & 1800.0 & $21.40 \pm 0.10$ & 1 \\
2009 Sep 3.92 & 1.47 & WHT-LIRIS & $K$ & 2592.0 & $18.92 \pm 0.20$ & 1 \\
2009 Sep 3.99 & 1.54 & GROND & $r^{\prime}$ & 738.0 & $21.54 \pm 0.05$ & 1 \\
2009 Sep 4.00 & 1.55 & NOT & $V$ & 900.0 & $21.67 \pm 0.11$ & 1 \\
2009 Sep 4.01 & 1.56 & NOT & $R$ & 900.0 & $21.40 \pm 0.11$ & 1 \\
2009 Sep 4.02 & 1.57 & NOT & $I$ & 900.0 & $20.72 \pm 0.11$ & 1 \\
2009 Sep 4.36 & 1.91 & UKIRT-WFCAM & $J$ & 1080.0 & $20.20 \pm 0.20$ & 1 \\
2009 Sep 4.36 & 1.91 & UKIRT-WFCAM & $K$ & 1080.0 & $18.90 \pm 0.25$ & 1 \\
2009 Sep 4.98 & 2.53 & GROND & $r^{\prime}$ & 738.0 & $22.01 \pm 0.07$ & 1 \\
2009 Sep 5.93 & 3.48 & Liverpool & $R$ & 1800.0 & $22.60 \pm 0.25$ & 1 \\
2009 Sep 6.37 & 3.97 & UVOT & $U$ & 10549.1 & $22.25 \pm 0.21$ & * \\
2009 Sep 8.98 & 6.53 & GROND & $r^{\prime}$ & 738.0 & $23.07 \pm 0.17$ & 1 \\
2009 Sep 10.37 & 7.97 & UVOT & $U$ & 9969.0 & $> 22.56$ & * \\
\enddata
\tablenotetext{a}{UT at beginning of exposure.}
\tablenotetext{b}{Time from midpoint of exposure to \fermi-GBM trigger.}
\tablenotetext{c}{Reported magnitudes have not been corrected for Galactic 
   extinction ($E(B-V) = 0.042$\,mag; \citealt{sfd98}).  Observations in the 
   the $r^{\prime}$ filter are reported on the AB magnitude system \citep{og83},
   while all other filters are referenced to Vega.}
\tablenotetext{d}{* -- This work; 1 -- \citet{psp+10}.}
\label{tab:0902opt}
\end{deluxetable}
%%%%%%%%%%%%%%%%%%%%%%%%%%%%%%%%%%%%%%%%%%%%%%%%%%%%%%%%%%%%%%%%%%%%%%%%%%%%%

\clearpage

%%%%%%%%%%%%%%%%%%%%%%%%%%%%%%%%%%%%%%%%%%%%%%%%%%%%%%%%%%%%%
\input{tab1.tex}
%\input{../Tables/0926aopt.tab.apj}
%%%%%%%%%%%%%%%%%%%%%%%%%%%%%%%%%%%%%%%%%%%%%%%%%%%%%%%%%%%%%

\end{document}